\def\tilde{\widetilde}
\def\bar{\overline}
\def\hat{\widehat}
\def\*{\star}
\def\[{\left[}
\def\]{\right]}
\def\({\left(}		
\def\){\right)}

\def\zb{{\bar{z} }}
\def\frac#1#2{{#1 \over #2}}
\def\inv#1{{1 \over #1}}
\def\half{{1 \over 2}}
\def\d{\partial}

\def\vev#1{\langle #1 \rangle}

\def\2pi{\hbox{$2\pi i$}}

\def\dsl{\raise.15ex\hbox{/}\kern-.57em\partial}
\def\Dsl{\,\raise.15ex\hbox{/}\mkern-.13.5mu D}

\def\ga{\gamma}		\def\Ga{\Gamma}

\def\al{\alpha}
\def\ep{\epsilon}
	\def\La{\Lambda}
\def\de{\delta}		\def\De{\Delta}
		
\def\sig{\sigma}	

\def\CG{{\cal G}}		
\def\CJ{{\cal J}}		
	\def\CN{{\cal N}}	\def\CO{{\cal O}}

\font\numbers=cmss12
\font\upright=cmu10 scaled\magstep1
\def\stroke{\vrule height8pt width0.4pt depth-0.1pt}
\def\topfleck{\vrule height8pt width0.5pt depth-5.9pt}
\def\botfleck{\vrule height2pt width0.5pt depth0.1pt}
\def\Zmath{\vcenter{\hbox{\numbers\rlap{\rlap{Z}\kern
		0.8pt\topfleck}\kern
		2.2pt \rlap Z\kern 6pt\botfleck\kern 1pt}}}
\def\Qmath{\vcenter{\hbox{\upright\rlap{\rlap{Q}\kern
                   3.8pt\stroke}\phantom{Q}}}}
\def\Nmath{\vcenter{\hbox{\upright\rlap{I}\kern 1.7pt N}}}
\def\Cmath{\vcenter{\hbox{\upright\rlap{\rlap{C}\kern
                   3.8pt\stroke}\phantom{C}}}}
\def\Rmath{\vcenter{\hbox{\upright\rlap{I}\kern 1.7pt R}}}
\def\Z{\ifmmode\Zmath\else$\Zmath$\fi}
\def\Q{\ifmmode\Qmath\else$\Qmath$\fi}
\def\N{\ifmmode\Nmath\else$\Nmath$\fi}
\def\C{\ifmmode\Cmath\else$\Cmath$\fi}
\def\R{\ifmmode\Rmath\else$\Rmath$\fi}
\def\cadremath#1{\vbox{\hrule\hbox{\vrule\kern8pt\vbox{\kern8pt
			\hbox{$\displaystyle #1$}\kern8pt} 
			\kern8pt\vrule}\hrule}}

\def\presentation{
\voffset -.50in   
\hoffset -.19in
\oddsidemargin 0in \evensidemargin 0in
\marginparwidth .75in \marginparsep 7pt \topmargin 0in
\headheight 12pt \headsep .25in
\footheight 18pt \footskip .35in
\textheight 9.5in \textwidth 6.5in
\columnsep 10pt \columnseprule 0pt }
\def\debut{ \begin{eqnarray} }
\def\fin{ \end{eqnarray} }
\def\non{ \nonumber }
\documentstyle[12pt]{article}
\presentation
\input epsf
\begin{document}
\rightline{SPhT-95-113}
\rightline{IHES/P/95/85}
\vskip 1cm
\centerline{\LARGE (Perturbed) Conformal Field Theory Applied To}
\bigskip
\centerline{\LARGE 2D Disordered Systems~: An Introduction.
\footnote[1]{Lectures presented at the '95 Cargese Summer School
on "Low dimensional application of quantum field theory".}}
\vskip 1cm
\vskip1cm
\centerline{\large  Denis Bernard
\footnote[2]{Member of the CNRS} }
\centerline{Service de Physique Th\'eorique de Saclay
\footnote[3]{\it Laboratoire de la Direction des Sciences de la
Mati\`ere du Commisariat \`a l'Energie Atomique.}}
\centerline{F-91191, Gif-sur-Yvette, France.}
\centerline{and}
\centerline{Institut des hautes \'etudes scientifiques,}
\centerline{F-91440, Bures-sur-Yvette, France.}
\vskip2cm
Abstract.\\
We describe applications of (perturbed) conformal field theories to 
two-dimensional disordered systems. We present various methods of study~: 

(i) {\it A direct method} in which we compute the explicit disorder 
dependence of the correlation functions for any sample of the disorder. 
This method seems to be specific to two dimensions.  The examples
we use are disordered versions of the Abelian and non-Abelian WZW models.
We show that the disordered WZW model over the Lie group $\CG$ at level $k$ 
is equivalent at large impurity density to the product of the WZW model
over the coset space $\CG^C/\CG$ at level $(-2h^v)$ times an arbitrary 
number of copies of the original WZW model.

(ii) {\it The supersymmetric method} is introduced using the random 
bond Ising model and the random Dirac theory as examples. In
particular, we show that the relevent algebra is the affine
$OSp(2N|2N)$ Lie superalgebra, an algebra with zero superdimension.

(iii) {\it The replica method} is introduced using the random phase sine-Gordon 
model as example. We describe particularities of its renormalization group flow. 

(iv) {\it A variationnal approach} is also presented using the
random phase sine-Gordon model as example. 

\vfill
\newpage
%
%
\section{Introduction.}

These lectures will probably look too elementary to
experts in disordered systems. It will perhaps be more
appropriate as an introduction to this field for conformal field theorists.

The plan of these notes is the following:

{\bf 1-Introduction}. We introduced the types of problems we will
be concerned with. We mention a few open problems;
the most challenging being the description of the quantum Hall
transition.

{\bf 2-A few basic techniques.} In this section we have gathered the 
basic tools which are usually used. This includes the supersymmetric
method and the replica method.

{\bf 3-The direct method.} There, we show that due to specific
properties of two-dimensional field theories, it is 
(sometimes) possible to compute directly the disorder dependence of
the correlation functions and then to average over the disorder.
Examples are provided by WZW models coupled to random vector potentials,
a particular case being the massless Dirac model coupled to
a random vector potential.
We discuss the Abelian and non-Abelian cases and show in both
cases that the disorder can be factorized.
At large impurity density, the disordered WZW model is conformally
invariant being equivalent to the product of the WZW model over the
non-compact space $\CG^C/\CG$ times an arbitrary number of copies
of the pure version of the original WZW model.

{\bf 4-The supersymmetric method.} We take the random Ising model 
as the simplest example for illustrating the supersymmetric
method. We show that it can be formulated as a Gross-Neveu type
model but over the Lie superalgebra $OSp(2N|2N)$, and we use
this formulation to recover its properties usually obtained
by the replica method. We discuss its connection with the
random Dirac theory: a model which has been introduced for
describing the quantum Hall transition. These examples illustrate 
the fact that non-unitrary conformal field theories based on
affine Lie superalgebras with zero superdimension provide 
candidates for critical theories for gaussian disordered systems.

{\bf 5-The replica method.} We take the random phase sine-Gordon model
as example. Its large distance behavior in the low temperature
phase is still controversial. We present the approach based
on the renormalization group and the symmetric replica trick. 
We also discuss a large $N$ version of this model.

{\bf 6-The variational method.} We again use the 
random phase sine-Gordon model to illustrate this method.
This approach consists in finding a good variational ansatz at
fixed disorder and then averaging. It provides exact lower bound
to the free energy.

Besides stochastic differential equations, cf. Polyakov's lecture 
in this volume, there are at least two classes of disordered 
problems which we can study.

i) The first ones concern statistical systems 
with random coupling constants, cf eg. \cite{LesHou}. Imagine having a
physical system with two kinds of degree of freedom
with two very different relaxation times. The 
degrees of freedom, called spin variables,
which have the shortest relaxation time will thermalyze
well before the other degrees of freedom, which we call
impurities. The impurities are also called quenched variables
because they are not in thermal
equilibrium with the spin variables.
We consider a large number of such physical systems
with different realizations of impurity configurations.
Since the spin variables are in thermal
equilibrium but not the impurities we are interested
in the average over the impurity configurations
of the free energy of the thermalyzed spin systems.

A classical example is provided by the Ising model
with random bond coupling constants. The thermalyzed
spin variables $\sig_i$, defined on the vertices
of a lattice, take two values $\sig_i=\pm$. The effect
of the impurities is represented by the randomness
of the interaction  $J_{ij}$ which couple
neighbourhood spins. At fixed disorder, its 
partition function is~:
\debut
Z[J] =  \sum_{\{\sig_i\}} \exp\({- \sum_{\vev{i,j} }
J_{ij} \sig_i \sig_j }\) \non
\fin
A probability distribution $P[J]$ is given to the random
coupling constants. The average thermal properties will
then be given by the average of the logarithm of the
partition function~: $\bar {\log Z[J]}$.

In the continuum limit and close to the critical point,
ie. in its scaling limit, it is known from prehistory 
that the Ising model is
described by a massive real Majorana fermion with mass
$\bar m\sim \tau = \frac{(T_C-T)}{T_C}$ with $T_C$ the critical temperature.
In presence of randomness in the bond interaction,
the mass becomes a random function of the space position.
Therefore, at fixed disorder, the action describing the
random bond Ising model in its continuous limit is~:
\begin{eqnarray}
S[m(x)]=\int \frac{d^2x}{4\pi}
\({ \bar \Psi i\dsl \Psi +   m (x) \bar \Psi \Psi }\)
\non
\end{eqnarray}
with $\Psi$ a Majorana fermion.
The random mass $m(x)$ is coupled to the energy operator
as it should be in order to represent the randomness of
the bond interaction.  This action is of the form~:
\debut
S[g(x)] = S_* + \int d^2 x~ g(x) \Phi(x)
\label{preums}
\fin
where $S_*$ is the action at the renormalization group
fixed point describing the pure system,
and $\Phi(x)$ a scalar field. As formulated in eq.(\ref{preums}),
the problem can also be interpreted as a random field problem.

The Ising model with random bond interactions provides
one of the simplest example of disordered system in 2d \cite{DoDo,Lu1,Shan}. We will
study it in details in the following (using the supersymmetric
method). Besides the disordered WZW models,
we will also study another example (using the replica
method)~: the random phase sine-Gordon model.
It has been introduced for describing a large variety
of  two dimensional disordered physical systems \cite{CaOs,VilFer,CuSha,Fish}.
Contrary to the random bond Ising model which is now well
understood, there is up to now no concensus
on the large distance behavior of the random phase
sine-Gordon model. We will present the various
suggestion which have been made concerning this behavior.


ii) The second class of problems deals with
particles moving in random potential, cf eg. \cite{LesHou}. The random potential
then represents the impurities that the particles, e.g.
the electrons, encountered while travelling in a metal.
The property of the medium will be
characterized by the  behavior of the electrons in this disordered
surrounding. For example, its conductivity properties will
depends on whether the averaged
 wave functions are localized or not. 
These problems are generically called localization problems.

We are thus interested in the averaged 
properties of  hamiltonians of the form~:
\debut
H = H_0 + V \non
\fin
where $H_0$ is a fixed hamiltonian,
and $V$ a random perturbation. A standard example is~:
$H_0$  a pure kinetic term, $H_0= -\d_x^2$ and $V=V(x)$ 
a potential representing the interaction with the underlying medium.

Quantities of interest would be e.g. the averaged density of states
which is given by the average of the trace of the Green function~:
\debut
\bar {\rho(E)} = \inv{\pi}\bar {Im ~tr\({\inv{H-E-i\ep}}\) } 
\label{dense}
\fin
More generally, we are interested in computing
 averages of products of Green functions. Indeed,
computing the conductivity tensor $\sig_{\mu\nu}$ 
using the Kubo formula requires computing
the average of the product of an advanced and a retarded
Green function.

Let me introduce a recent localization/delocalization
problem (which actually was the motivation for these lectures).
It concerns the transition between the plateaux of the
integer quantum Hall effect, cf \cite{Huck,Zirn} and references
therein. Without going into any details,
we recall that the quantum
Hall effect is characterized by the quantization of the transverse
Hall resistance $R_{xy}$ to inverse integer values in unit of $(h/e^2)$.
This integer quantization remains valid when the magnetic field 
varies on some domain, which are called plateaux. 
On these plateaux the longitudinal resistance $R_{xx}$
vanishes, but it becomes non-zero in the regime separating
two plateaux. The values of the resistance $R_{xy}$ and
$R_{xx}$ as a function of the magnetic field is represented
in Fig.1. 

$${\epsfxsize=6.5 truecm \epsfbox{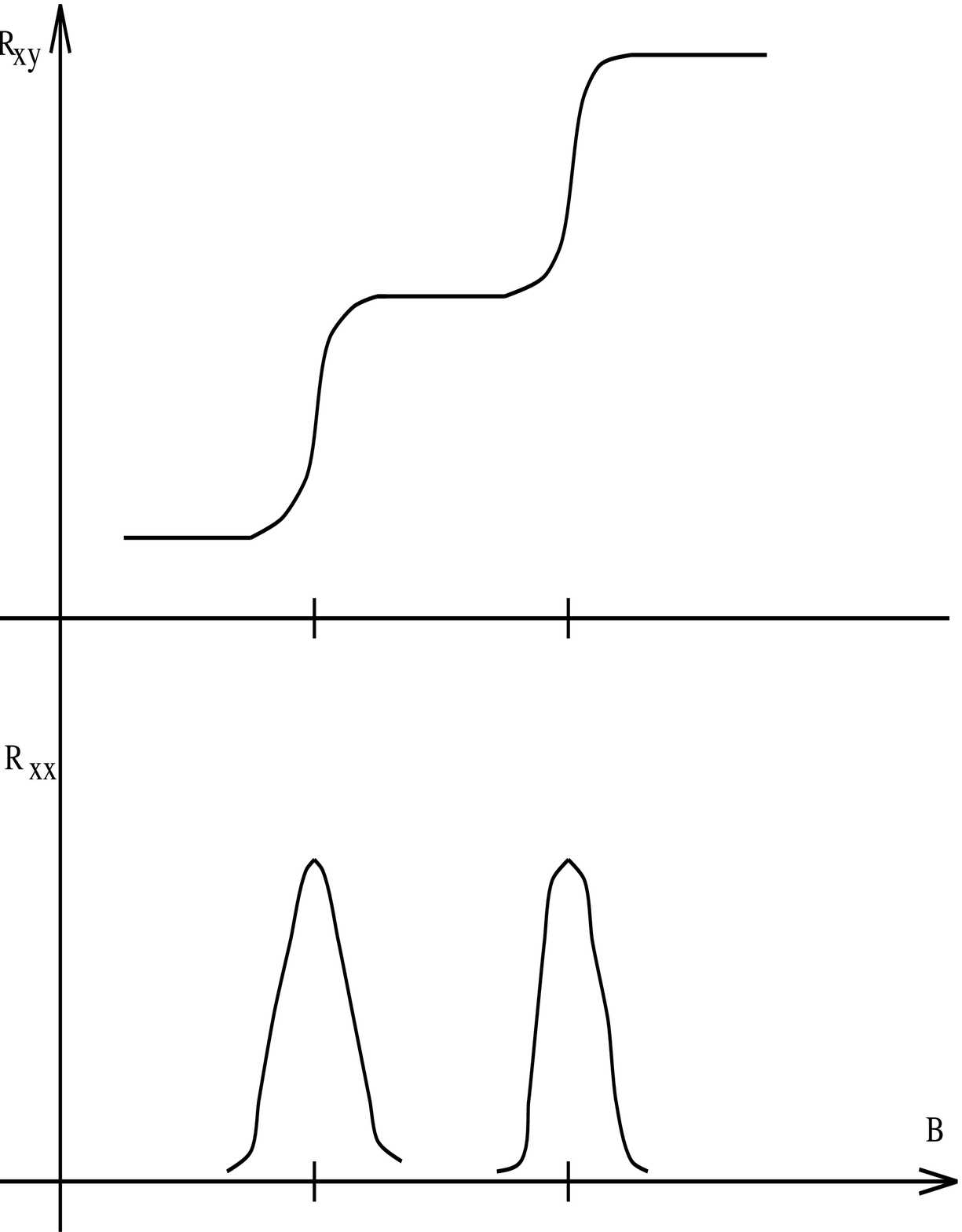}}$$
\centerline{Fig.1 : The transverse and
longitudinal Hall resistances $R_{xy}$ and $R_{xx}$}
\centerline{as a function of the magnetic field $B$.} 
\vskip 1.5 truecm

The integer quantum Hall
effect (IQHE) is correctly described by considering
a 2d gas of non-interacting  electrons  subject to a
transverse magnetic field. As is well known, the energy
spectrum is then given by the Landau levels, which
are highly degenerate.
In a pure system without any impurities, 
the Hall resistance $R_{xy}$ will
be a step-function as a function of the Fermi level.
The plateaux  occur when the Fermi level of the
electrons gas is between two Landau levels. 

However, since the degeneracy of a landau level is
field dependent, (this degeneracy is $\phi/\phi_0$ where
$\phi$ is the magnetic flux through the system and
$\phi_0=h/e$ the flux quantum), the Hall resistance
in a pure system is a linear function of the magnetic field $B$.
More precisely~:
\debut
R_{xy}^{pure}= (\frac{h}{e^2}) \frac{\phi}{N_e\phi_0}
= (\frac{h}{e^2}) \inv{\nu} \non
\fin 
where $N_e$ is the number of electrons in the system and
$\nu= \frac{N_e\phi_0}{\phi}$ is called the filling factor.
The explanation of the  plateaux for $R_{xy}$ 
as a function of
the magnetic field $B$ and not of the Fermi level
requires considering the effect of the impurities \cite{Laug,Halp}.
The impurity potential splits the degeneracy of the Landau
levels, which now form energy bands. 
The states whose energy is in the border of
these bands are localized while those which are in the center
of the bands are delocalized. 
The localized states then
serve as reservoir for the electrons which stabilized the
Fermi level between two Landau bands while the magnetic
field varies. This stabililization of the Fermi level
give rise to the plateaux.
It is expected that the delocalized states are present
only at the center of the band; that is there is not a band of 
delocalized states. The presence of delocalized
states is necessary for having a non-zero conductivity.

$${\epsfxsize= 6.5 truecm \epsfbox{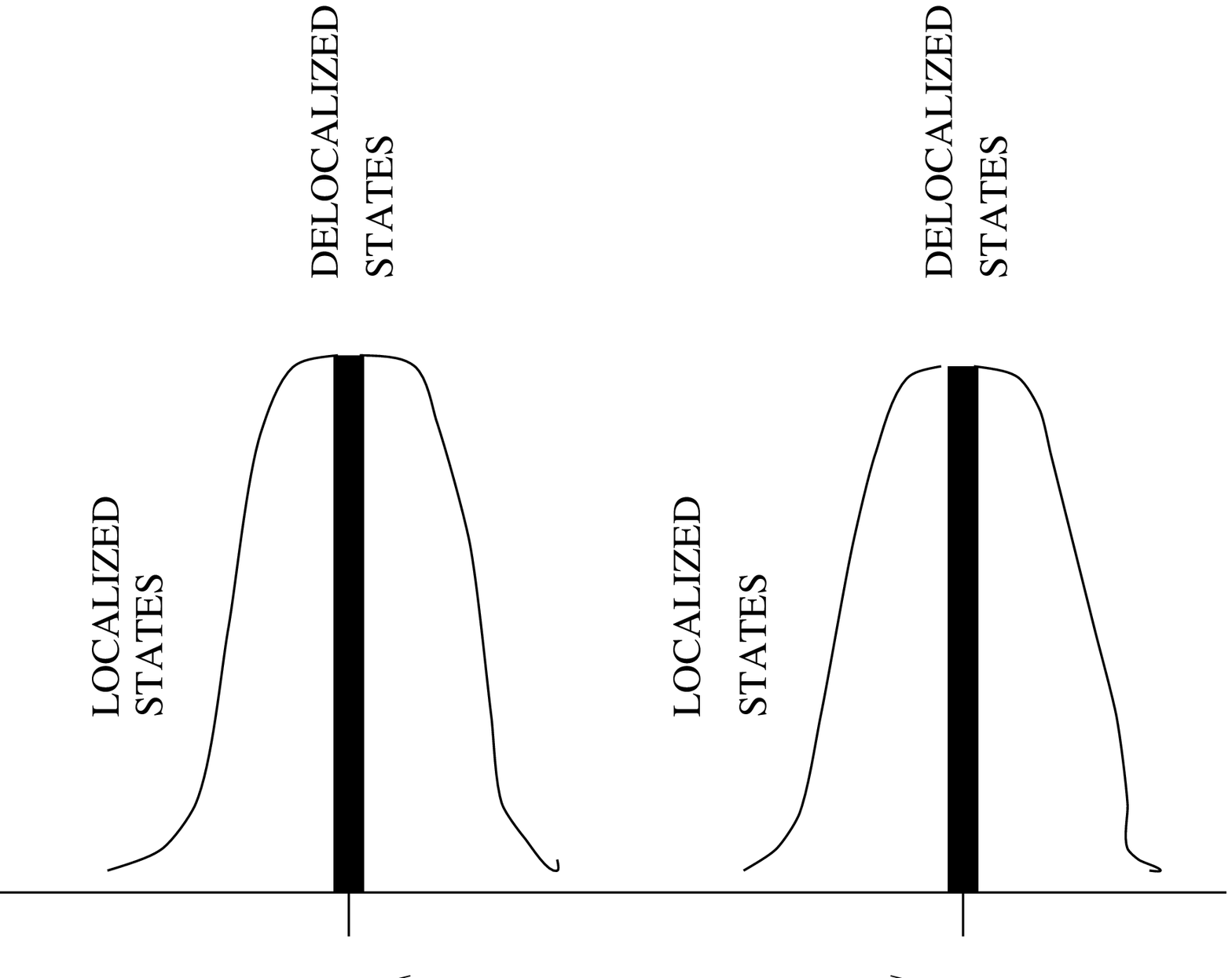}}$$
\centerline{Fig.2 : The density of states in two consecutive Landau bands.}
\vskip 1.5 truecm

Between two plateaux the Fermi level is (almost) in the 
middle of a Landau band. Since the delocalized states
are in the middle of this band, the transition between
two plateaux of the integer quantum Hall effect is thus
a transition between localized and delocalized states.
(De)localized states are charactized by a (in)finite
localization length. Near this transition, this
 localization length $\xi(E)$ behaves as follows~:
\debut
\xi(E) \sim |E-E_c|^{-\nu} \label{loca}
\fin
with $E_c$ the band center energy. 
The density of states is non-singular at the transition point \cite{BreIt}.
Various mesurements, as well as  numerical
simulations, gave access to the exposant $\nu$~:
$\nu\sim 2.3 \pm 0.1\sim 7/3$ \cite{Kochetal,Weietal,Huck}.
The behavior (\ref{loca}) has to be compared to the
standard Anderson localization in two dimensions which
predicts that any tiny amount of disorder will localize
all the states.  

What is the field theory describing the quantum Hall transition
is still an open question, cf ref.\cite{Zirn} for a recent
discussion. The problem consists in determining
the  appropriate infrared  fixed point, and therefore seems
to be a ``simple" exercise in conformal field theory.
Unfortunalely, this infrared fixed point is a strong coupling
problem in all the models introduced so far; it is therefore difficult to attack.
A sigma model approach was introduced in ref.\cite{Pruisk,Weism}.
A very elegant network model was developped in ref.\cite{ChakCodd}.
A supersymmetric spin chain, which arises as an anisotropic
limit of the network model, was introduced in ref.\cite{Zirn}.
A model closely related to the random bond 
Ising model, namely the random Dirac model,
has also been proposed for describing the IQHE
localization/delocalization transition \cite{Luetal}.
We will present it in section 4.

\section{ A few basic techniques.}

These problems can be analyzed with similar techniques. 
The two standard ways of studying them use either
the replica trick, cf eg \cite{LesHou} or the supersymmetic method \cite{Efet}.
In this section we present a very brief introduction to
these methods. (The similarity of the techniques
used to study disordered statistical model and localization
problems  is apparent at a first naive level, but more refined
studies of both subjects reveal their differences.)

As it will become clear below, the supersymmetric method applies
to system which are gaussian at fixed disorder, while
the replica trick has a larger domain
of applicability since it is not restricted to free theory. 
Perturbatively the replica and supersymmetic methods are equivalent.
However, a few non-perturbative results have been obtained using
the supersymmetric method.
Clearly, once these disordered problems have been reformulated in a field
theoretical languages, all the standard field theoretical
methods (e.g renormalization group,
mean field approximation, variational approximation,
large $N$ developments, etc...) are avalable.

In the following sections we will illustrate these methods
on two examples~: the random bond
Ising model and the random phase sine-Gordon model.

\bigskip
$\bullet)$ {\bf The supersymmetric method.}

The supersymmetric method \cite{Efet} applies to models which are
gaussian at fixed disorder. We will present it in the random
Hamiltonian problem. As above let us consider a
random hermitian hamiltonian 
\debut
H=H_0+V, \non
\fin
 where $H_0$ is a fixed
hamiltonian and $V$ a random potential. 
We are interested in computing averages of product of
advanced or retarded Green functions~:
\debut
\bar { \({\inv{H_E-i\ep}}\)\({\inv{H_{E'}+i\ep}}\) \cdots }\non
\fin
with $\ep\to 0^+$. We have set $H_E=H-E$.
The supersymmetric method is based on a gaussian
integral representation of the Green functions. 
E.g.:
\debut
\inv{H_E-i\ep} = \inv{Det[H_E-i\ep]}\int d\psi~
 \psi\psi^* ~\exp\({ \pm i \psi^* (H_E-i\ep) \psi }\)
\label{repfer}
\fin
with $\psi$ complex grassmanian variables.
In eq.(\ref{repfer}), we have the freedom to choose the
$\pm$ sign. There are no convergence problem in this
fermionic integral. To be able to average over the disorder,
we need to re-exponentiate the determinant $Det[H_E-i\ep]$.
The inverse of this determinant can be written
as a gaussian integral over complex bosonic variables~:
\debut
\inv{Det[H_E-i\ep]} = \int d\phi ~exp\({ -i \phi^*(H_E-i\ep)\phi }\)
\label{repbos}
\fin
The prefactor $(-i)$ in the exponential is choosen by
requiring the convergence of the gaussian integral.
Gathering these gaussian integrals, we obtain the supersymmetric
representation of the Green functions~:
\debut
\({\inv{H_E-i\ep}}\) &=& \int  d\psi d\phi~ \psi\psi^*~
\exp\({ -i \psi^* (H_E-i\ep) \psi -i \phi^*(H_E-i\ep)\phi}\)
\label{repsusy}\\
\({\inv{H_E-i\ep}}\)\({\inv{H_{E'}+i\ep}}\)
&=& \int d\psi d\phi~\psi_R\psi_R^*\psi_A\psi_A^* 
\exp \({ -i \psi_R^* (H_{E}-i\ep) \psi_R -i \phi_R^*(H_{E}-i\ep)\phi_R 
 }\right.\non\\
&~& ~~~~~~~~~~~~~~~~~~~~ 
\left.{ -i \psi_A^* (H_{E'}+i\ep) \psi_A 
+i \phi_A^*(H_{E'}+i\ep)\phi_A }\)\non\\
 ... etc ...\non
\fin
In the equation representing product of two Green
functions,  the indices $A,R$
refers to the advanced or retarded Green functions.
Note that due to the convergence requirement, the
sign in front of the $(i\ep)$ factor differs in the advanced
and retarded sector.
More generally, products of $N$ Green functions are represented by 
integrals over
$N$ couples of superpartners $(\psi,\phi)$ with appropriate choice
of the sign factor according to the convergence condition.

In the form (\ref{repsusy}), the average over the disorder can be
easily done. Assume for simplicity that the random potential has
a gaussian distribution~:
\debut
P[V] = \exp{\({ -\inv{2\sig} tr(V^2)}\) } dV \non
\fin
Consider for example the average of one Green function.
The term coupled to the random potential is
$\inv{2\sig}tr(V^2)+itr(V(\phi\phi^*-\psi\psi^*))$.
After integration over $V$,
the averaged Green functions becomes~:
\debut
\bar {\({\inv{H-i\ep}}\)} =\int d\psi d\phi~ \psi\psi^* 
\exp\({- S_{eff} }\) \label{susyeff}
\fin
with 
\debut
 S_{eff} = 
i \psi^* (H_0-E-i\ep) \psi +i \phi^*(H_0-E-i\ep)\phi 
-\frac{\sig}{2} tr\({\phi\phi^*-\psi\psi^*}\)^2\non
\fin
The bosonic and fermionic sectors are coupled by the disorder.
This effective action is easily generalized when considering
products of  higher number of Green functions.

Notice that if $H$ is a quantum hamiltonian for a physical
system in $d$ spacial dimensions, ie. a $(d+1)$ system,
than the supersymmetric action describes a d dimensional 
quantum field theory. In particular, since the quantum Hall effect
is a $(2+1)$ system its localization/delocalization transition
would be described by a 2d field theory. 

\bigskip
$\bullet)$ {\bf The replica method.}

Let us now recall briefly the starting point of 
the replica method, cf \cite{LesHou}. We are interested in
computing the average of the free energy, or of
correlation functions.  Since the free energy is 
proportional to the logarithm of the partition function $Z[J]$,
we have to compute the average of $\log Z[J]$.
The replica method is based on the identity~:
\debut
\log Z[J] = \lim_{n\to 0} \frac{Z^n[J]-1}{n}
= \frac{d}{dn} Z^n[J] \Big\vert_{n=0} \label{logz}
\fin
In other words, we compute the quenched averages of one
disordered system by introducing
$n$ copies of this system, in all of which the random
variables take the same values, and then computing the 
averaged partition function of this new system at $n=0$.

This method leads immediatly to a criteria selecting
the systems in which the disorder is relevent. This
criteria is called the Harris criteria. 

Let us describe a disordered system in the continuum limit by
the action~:
\debut
S[g(x)] = S_* + \sum_a \int d^2 x~ g^a(x) \Phi_a(x) 
\label{actgx}
\fin
where $S_*$ is the action at the renormalization group
fixed point describing the pure system at its critical point,
and $\Phi_a(x)$ are scalar fields. When the coupling constant
$g^a(x)$ are independent of $x$ and not random, the action
(\ref{actgx}) describes the behavior of the statistical
model near the critical point in its scaling limit. 
In this case the field $\Phi_a(x)$ is relevent, and therefore
will influence the large distance behavior of the pure system,
if its scaling dimension is less than two~: $\dim(\Phi_a)<2$.
This is changed when disorder is present.

Assume for simplicity that the random variables $g^a(x)$
are gaussian variables with one and two-point functions~:
\debut
\bar {g^a(x)} = 0 \quad and \quad
\bar{ g^a(x) g^b(y) } =   \delta^{ab}\sig_a \delta(x-y)
\label{mesgau}
\fin
Before averaging over the disorder, the partition function
is $Z[g(x)]=\int DX ~\exp(-S[g(x)])$ and the $n$ replicated
one becomes~:
\debut
Z^n[g(x)] = \int \prod_r DX_r~
\exp\({ - \sum_r S_*^r - \sum_a \int d^2 x~ g^a(x)\sum_r\Phi^r_a(x)}\)
\non
\fin
where the index $r$ runs from $1$ to $n$, 
$S_*^r$ is the action for the $r^{th}$ replic and
$\Phi^r_a(x)$ is the field $\Phi_a(x)$ in the
$r^{th}$ copy. To have introduced the replicated copies allows us to 
easily average over the disorder. Assuming the gaussian measure
(\ref{mesgau}) for $g^a(x)$, we obtain~:
\debut
\bar {Z^n[g(x)]} =  \int \prod_r DX_r~
\exp\({ - S_{eff} }\)
\label{replic}
\fin
with
\debut
S_{eff}= \sum_r S_*^r + \half\sum_a \sig_a  \int d^2 x~
\sum_{r,s} \Phi^r_a(x)\Phi^s_a(x) \label{Srepl}
\fin
The replica are now coupled by the disorder. 
The field perturbing the uncorrelated action $\sum_r S_*^r$
is quadratic in terms of the $\Phi_a(x)$'s. 
It will be relevent if $2\dim\({\Phi_a(x)}\)<2$.  (A more careful
study is needed to analyse the short distance singularity
in $\Phi^r_a(x)\Phi^s_a(x)$ when $r=s$).
Hence~:
\debut
\dim{\Phi_a(x)} < 1  \Rightarrow  relevent~~disorder. \non
\fin
This is the Harris criteria in two dimensions. Note that for
a field to be relevent in a quenched system its dimension
has to be half what is required in the pure system.
If these dimensions depend on parameters the Harris criteria
provides a first glance to the phase diagram.

The quenched correlation functions are defined by~:
\debut
\bar {\vev{\CO(x)\cdots}\vev{\cdots}} =
\bar {\({ \inv{Z[g(x)]} \int DX~ e^{-S[g(x)]}~ \CO(x)\cdots}\)\cdots}
\non
\fin
The replica method can also be used to compute the averaged
correlation functions. It provides a way to re-exponentiate
the inverse partition function. The rules are the following~:
\debut
\bar {\vev{\CO(x)}} &=& \vev{\CO^r(x)} \vert_{n=0} \non\\
\bar {\vev{\CO(x)\CO(y)}} &=& \vev{\CO^r(x)\CO^r(y)} \vert_{n=0}
\label{correpl}\\
\bar {\vev{\CO(x)}\vev{\CO(y)}} &=& \vev{\CO^r(x)\CO^s(y)} \vert_{n=0},
\quad r\not= s\non\\
  ...&etc&....\non
\fin
In particular, the average of the connected Green functions
are represented by the correlation function of 
$\inv{n}\sum_r \CO^r(x)$.
The simplest way to prove eq.(\ref{correpl})
 consists in introducing sources in the quenched
partition function.

\section{The direct method~: the random vector potential model.}

\def\Asl{\raise.15ex\hbox{/}\kern-.67em A}
\def\zb{ {\bar z} }
\bigskip

As a warm up we introduce
a very simple gaussian model, which evidently can be
completely solved. (We will later use a few of its properities
in our study of the more interesting random phase sine-Gordon model.)

In two dimensions it is sometimes possible to compute
exactly field theory partition functions and 
correlation functions with sources. 
In this section, we also described how this property
can be applied to very particular disordered 2d
systems. The example we choose consists in a non-Abelian 
generalization of the gaussian model based on the WZW models.
It provides one of the rare disordered models which can
be solved directly without relying on the supersymmetric
or replica method. This solution relies on the fact that
the disorder dependence can be factorized.

\bigskip
$\bullet$) {\bf The gaussian model.}

The bosonic form of its action  is defined to be~:
\debut
S_0[\Phi,A]&=& \int\frac{d^2x}{4\pi}\({
\half (\d_\mu\Phi)^2 + iA_\mu \ep_{\mu\nu}\d_\nu\Phi }\),\non\\
&=& \int\frac{d^2x}{\pi}\({\half\d_z\Phi\d_\zb\Phi +
i(A_\zb\d\Phi+A_z\d_\zb\Phi) }\).
\label{act0}
\fin
where $A_\mu$ is a random quenched variable with
measure~:
\debut
P[A]= \exp\[{ - \inv{2\sig}\int~\frac{d^2x}{4\pi}~ A_\mu A_\mu }\] 
= \exp\[{ - \inv{\sig}\int~\frac{d^2x}{\pi}~ A_z A_\zb }\]
\label{mesA1}
\fin
In eq.(\ref{act0}) we introduced the complex coordinate 
$z=x+iy$, $\bar z=x-iy$ on the plane.

Since it is a gaussian model, it can be easily solved
without using the replica trick or the supersymmetric
formulation, but by directly computing the
quenched averages.
Let us introduce the Hodge decomposition of $A_\mu$~:
$A_\mu=\d_\mu\xi + \ep_{\mu\nu}\d_\nu\eta$. 
The fields $\xi$ and $\eta$ decouple in the measure (\ref{mesA1})~:
\begin{eqnarray} 
P[A]=P[\xi;\eta] = 
  \exp\[{ - \inv{2\sig}\int~\frac{d^2x}{4\pi}~
 \({(\d_\mu\xi)^2 + (\d_\mu\eta)^2}\) }\] \label{meseta}
\end{eqnarray}
The action $S_0[\Phi,A]$ is independent of $\xi$ and
therefore the field $\xi$ is irrelevent. This fact was expected
since the field $\xi$ represents a pure gauge whereas the physically
relevent quantity is the field strengh $F=\ep_{\mu\nu}\d_\mu A_\nu
= (\d_\mu\d_\mu) \eta$. Moreover, the field $\eta$ can be 
absorbed into a translation of $\Phi$~:
\debut
S_0[\Phi,A_\mu=\ep_{\mu\nu}\d_\nu\eta]=
S_0[\Phi+i\eta,A=0] + \int\frac{d^2x}{4 \pi} 
\half(\d_\mu\eta)^2.
\label{anu1}
\fin

The fact that the field $\eta$ can be 
absorbed into a shift of $\Phi$ does not mean that the quenched
correlation functions are identical to those in the pure 
system. Using eq.(\ref{anu1}), we have
\debut
\vev{ \prod_n e^{i\al_n\Phi(x_n)} }_{A_\mu=\ep_{\mu\nu}\d_\nu\eta}
= e^{\sum_n \al_n\eta(x_n)}~~ 
\vev{ \prod_n e^{i\al_n\Phi(x_n)} }_{A=0}.\non
\fin

It factorizes into the product of the correlation functions
in the pure system times a simple function of the impurities.
However, the average of this function is not irrelevent since
the variables $\eta$ have long-range correlations~:
${\bar {\eta(x)\eta(y)}}= -\pi\sig (\d_z\d_\zb)^{-1}(x,y)
= -\sig \log|x-y|^2$. In particular, it changes the values
of the critical exponents. 

Averages of products of correlation functions can then be computed .
One has~:
\begin{eqnarray} 
{\bar {\vev{\prod_n e^{i\al_n\Phi(x_n)} }_A \cdots
\vev{\prod_m e^{i\beta_m\Phi(y_m)} }_A}}  &=&
\vev{\vev{e^{\sum_n\al_n\eta(x_n)+\cdots+
\sum_m\beta_m\eta(y_m)}}}~\times \non\\
& ~& \times~\vev{\prod_n e^{i\al_n\Phi(x_n)} }_{A=0}~\cdots~
\vev{\prod_m e^{i\beta_n\Phi(y_m)} }_{A=0}
\non
\end{eqnarray}
where $\vev{\vev{\cdots}}$ refers to the $\eta$-correlation 
functions with the free field measure (\ref{meseta}).
All the correlators are therefore given by gaussian integral.
We get~:
\debut
{\bar {\vev{\prod_n e^{i\al_n\Phi(x_n)} }_A \cdots
\vev{\prod_m e^{i\beta_m\Phi(y_m)} }_A}} &= & ~~~~~~~~~
\prod_{n,m}{\vert{x_n-y_m}\vert}^{-2\sig \al_n\beta_m}\times\label{corrab}\\
 &~&\times ~~\prod_{n<n'}{\vert{x_n-x_{n'}}\vert}^{2\al_n\al_{n'}}\cdots
\prod_{m<m'}{\vert{y_m-y_{m'}}\vert}^{2\beta_m\beta_{m'}}\non
\fin
Clearly, conformal invariance is unbroken in the random abelian case. 
The dimensions $\De$ of the vertex
operators $\exp(i\al\Phi(x))$ in the quenched theory are~:
\debut
\De_{quenched}^\al = \De_{pure}^\al - \sig \al^2. \non
\fin
It is interesting to look at the connected Green functions of
the current $J_z=i\d\Phi$. It turns out that all
these quenched connected correlations are chiral; in particular,
they are holomorphic and only depend on $z=x+iy$.
For example, we have~:
\debut
\bar{\vev{i\d\Phi(z)}} &=& 0,\non\\
\bar{\vev{i\d\Phi(z)i\d\Phi(w)}_{conn.}} &=& 
2\pi\({\d\inv{\bar \d}}\)_{z,w}= \inv{(z-w)^2}\non
\fin
More generally, one  verifies that~:
\debut
\bar{\vev{i\d\Phi(z_1)\cdots i\d\Phi(z_P)}_{conn.}} &=& 
\vev{i\d\Phi(z_1)\cdots i\d\Phi(z_P)}_{A=0}.\label{connect}
\fin
In other words, the averages of these connected correlation
functions are unaffected by the disorder. This properity is not true
for the average of disconnected correlation functions of  $J_z=i\d\Phi$.

A more complete derivation of these Ward identities which follows
from a $U(1)$ symmetry is presented in the section devoted to
the study of the random phase sine-Gordon model.

It is an interesting exercise to decipher the operator product
algebra of these models.

\bigskip
$\bullet)$ {\bf The non-Abelian case~: factorization.}

We now turn to the non-Abelian case which provides
an exceptional quenched model which can be solved 
directly, without using any replica or supersymmetric method.
This is done by factorizing the disorder dependence.
We derive it directly from
the definition of the quenched correlation functions.
This result was also found in ref.\cite{Murp}  using the 
supersymmetric method, (it appears while we were
preparing these notes). 

The non-Abelian model appears naturally when  
generalizing the previous model, after fermionization,
to a $N$-component model.
Indeed, using  standard bosonization/fermionization rules, 
$\bar \Psi\ga_\mu\Psi = \ep_{\mu\nu}
\d_\nu\Phi$, and $\bar \Psi i\dsl \Psi = \half (\d_\mu\Phi)^2$,
we can rewrite the action  (\ref{act0}) in an 
equivalent fermionic form~:
\debut
S_0= \int\frac{d^2x}{\pi}\({
\bar \Psi(i\dsl+\Asl)\Psi}\).\non
\fin
It describes a Dirac fermion coupled to a random magnetic field.

The non-Abelian generalization describes $N$ massless 
Dirac fermions minimally coupled to a random
non-abelian gauge field \cite{Neras,Be,Murp}. Let us introduce 
the components of the fermions $(\bar \psi^j_+,\psi^j_+)$ and
$(\bar \psi_{-;j},\psi_{-;j})$, $j=1,\cdots,N$. The action
is  defined as~:
\debut
S[A]=\int \frac{d^2x}{\pi}
\({\psi_{-;j}\( \d_\zb \de^j_k + A^j_{\zb,k}\)\psi^k_+ +
\bar \psi_{-;j}\( \d_z \de^j_k + A^j_{z,k}\)\bar \psi^k_+ }\)
\label{action}
\fin
where $A^j_{z,k}=i\sum_a A^a_z (t^a)^j_k$, with $(A^a_z)^*=A_\zb^a$,
is the gauge field.
Here the hermitian matrices $t^a$ form the $N$-dimensional representation
of $SU(N)$. We denote by $f^{abc}$ the $SU(N)$ structure constants~:
$[t^a,t^b]=if^{abc}t^c$. The Dirac fermions take values in this
$N$-dimensional representation.
The gauge field is assumed to be a quenched variable with 
the gaussian measure~:
\debut
P[A]=\exp\[{-\inv{\sig}\int \frac{d^2x}{\pi}\sum_a A^a_z A^a_\zb}\]
\label{mesnona}
\fin

At fixed disorder the partition function is~:
\debut
Z[A] = e^{W[A]}= \int D\psi e^{-S[A]} = Det[i\dsl+\Asl] \label{partnona}
\fin
It can be expressed in terms of the WZW action \cite{PW}, see below 
eq.(\ref{partwzw}).

The way to factorize the disorder consists in implementing
a chiral gauge transformation. So we parametrized the vector
potential by an element $G$ of the complexified  group $SU(N)^C$
as follows~:
\debut
A_\zb = G^{-1}~(\d_\zb ~G) = - A_z^* \label{paranona}
\fin
This is always possible on the sphere provided the connexion $A$ is regular
enough. This parametrization is unique up to left multiplication
by a constant group element. Notice that the number of degree
of freedom is preserved by this transformation~: in the $A$ or
$G$ parametrization there are $2dimSU(N)$ degree of freedom.
We then have $\psi_-(\d_\zb+ A_\zb)\psi_+= (\psi_-G^{-1})
\d_\zb (G\psi_+)$ and $\bar \psi_-(\d_z+ A_z)\bar \psi_+= 
(\bar \psi_-G^*) \d_z (G^{-1\;*}\bar \psi_+)$. 
We can therefore absorbed the dependence on $A$ by transforming
the fermions as~:
\debut
\hat \psi_- = \psi_-G^{-1} \quad &;& \quad
 \hat {\bar \psi_-}  = \bar \psi_-G^* \non\\
\hat \psi_+ = G\psi_+ \quad &;& \quad
 \hat {\bar \psi_+} = G^{-1\;*}\bar \psi_+ \label{chiralnona}
\fin
This is the chiral gauge transformation. It maps the action (\ref{action}) into
the free Dirac action for the gauge transformed fermions.
We denote it $S_{free}[\hat \psi]$. The Jacobian for
the transformation $\psi \to \hat \psi$ is non-trivial but
given by the chiral anomaly~:
\debut
\Big\vert{\frac{D\psi}{D\hat \psi}}\Big\vert=
\frac{ Det[i\dsl+\Asl]}{Det[i\dsl]} = Z[A] \non
\fin
The extra factor $Det[i\dsl]$ is irrelevent when computing the
quenched correlation function but relevent in the evaluation of
the conformal anomaly.
The crucial point is that this Jacobian is equal to the partition function
at fixed disorder. Therefore, the partition and the Jacobian
simplifies when computing the correlation functions at fixed
disorder, and we get~:
\debut
\vev{ \psi_- \cdots }_A &=& \inv{Z[A]}
\int D\psi~ e^{-S[A]}~ \psi_- \cdots  \non\\
&=& \int D\hat \psi~ e^{-S_{free}[\hat \psi]}~ (\hat \psi_- G)\cdots
\label{factonona}
\fin
This is the announced factorization property~: the correlation
functions at fixed disorder factorize into the product of
a correlation function in the free Dirac theory times an 
explicitely known functional of the vector potential.

We now have to average over the disorder. For this it is
convenient, if not necessary, to change variable from
$A$ to $G$. The Jacobian is known, see eg. \cite{GaKu}~:
\debut
DA = DG ~\exp\({~ 2h^v S_{wzw}(GG^*)~}\) \label{jacob}
\fin
where $DG$ is the Haar measure on $SU(N)^C$, 
$S_{wzw}(GG^*)$ is the WZW action and $h^v$ is the
dual Coxeter number, equals to $N$ is the $SU(N)$ case.
Since only the product $GG^*$ enters into the Jacobian, 
we decompose $SU(N)^C$ as $(SU(N)^C/SU(N))\times SU(N)$.
For $G$ this means~:
\debut
G= H~ U \quad with\quad H \in SU(N)^C/SU(N),~~
U\in SU(N) \non
\fin
The Haar measure $DG$ factorizes on the Haar measure
on $(SU(N)^C/SU(N))$ and $SU(N)$. Thus, we finally obtain~:
\debut
DA = DU DH \exp\({ 2h^v S_{wzw}(HH^*)~}\) \label{jacobbis}
\fin
In eq. (\ref{jacobbis}) the part related to the compact space $SU(N)$ only
involves the Haar measure, while the part related to the non-compact
symmetric space $(SU(N)^C/SU(N))$ involves the WZW action.
Although the WZW action seems to appears with a `wrong' sign
in eq.(\ref{jacobbis}),
the non-compact sigma model is well defined since the metric
on $(SU(N)^C/SU(N))$ is negative definite.

For an arbitrary value of the disorder strengh $\sig$ we also have
to include the factor (\ref{mesnona}) into the measure for $A$.
This factor couples $U$ and $H$.
But things simplify at $\sig=\infty$. For this value of the
coupling constant, we have~:
\debut
P_{\sig=\infty}[A]~DA = DU DH \exp\({ 2h^v S_{wzw}(HH^*)~}\) \label{simple}
\fin
The compact and non-compact sectors are now independent.
Moreover, the field U of the compact sector completely 
decouples from the quenched correlation functions 
of the local spinless neutral operators as e.g.
$(\psi_-\bar \psi_+)$. Indeed, we have~:
\debut
\bar {\vev{ (\psi_-\bar \psi_+)\cdots}_A\cdots }_{~~\sig=\infty}
=\int DHD\hat\psi~ e^{2h^vS_{wzw}(HH^*)-S_{free}[\hat\psi]}~
(\hat\psi_-~HH^*~\bar {\hat\psi^+})\cdots \label{bignona}
\fin
Notice that a necessary condition for the cancelation
of the $U$-dependence in eq.(\ref{bignona}) is to consider
$SU(N)$ scalar operators. Ie. we have to sum over the
color indices of the fermions $\psi_-$ and $\bar \psi_+$ in
order to multiply $U$ and $U^*$ to get $UU^*=1$.

The free Dirac action plus the non-compact sigma model
are conformally invariant. This is clear for the 
free massless Dirac theory.
The sigma-model on the non-compact
space $(SU(N)^C/SU(N))$ defined by the WZW action  
was studied in ref.\cite{GaKu} in connexion
with path integral construction of the coset models.
There it is was shown that it is a non-unitary conformal field theory  
which carries a representation of the affine
Kac-Moody algebra of negative level $\tilde k = -2h^v$. 
Therefore, the Virasoro central charge of the non-compact
sector is~:
\debut
C = \frac{(-2h^v) dim~su(N)}{(-2h^v) + h^v} = 2 dimSU(N)=2 (N^2-1)\label{cvircoset}
\fin
The scaling dimension of the primary fields $\phi_R$ which belong to
a representation $R$ are given as usual by~:
\debut
dim[\phi_R] = 2\frac{ Cas(R)}{2( (-2h^v)+h^v)}= -2 \frac{Cas(R)}{2h^v}
\label{confcoset}
\fin
with $Cas(R)$ the Casimir operator in $R$. The negative value of
this dimension is an echo of the non-unitary character 
of the theory.

\def\bbbox{\begin{picture}(3,3)(-3,-3)
\put(-3,-3){\framebox(3,3)} \end{picture}}
As an application of eq.(\ref{bignona}) we can derive the scaling
dimension of the operator $(\psi_-\bar \psi_+)$ in
the quenched correlation functions. It is the sum of the scaling
dimensions of the operator $(\hat \psi_-\bar {\hat \psi_+})$
in the free Dirac theory plus the one of the operator
$(HH^*)$ in the vector representation $(\bbbox)$ in the non-compact
WZW model. Explicitely~:
\debut
dim_{\sig=\infty} [\bar {(\psi_-\bar \psi_+)}]
= 1 - 2 \frac{Cas(\bbbox)}{(-2N)} = 1 -\frac{N^2-1}{N^2} \non
\fin
This agrees with ref.\cite{Neras,Murp}.

In summary, we have shown that at $\sig=\infty$ we have the
following equivalence, valid in the local spinless neutral sector~:
\debut
{\bar {Dirac_A}}_{~~\sig=\infty} = (Free~ Dirac)^\CN
\times (SU(N)^C/SU(N))_{\tilde k=-2h^v} \non
\fin
with $\CN$ the number of correlations
we are averaging.
This equivalence is a direct consequence of the
factorization property eq.(\ref{simple},\ref{bignona})
of the random vector potential model.

\bigskip
$\bullet$) {\bf Quenched current correlation functions.} 

The previous factorization property is really useful
only at infinite coupling $\sig$. Here we describe how
averages of current correlation functions can be computed
for any value of $\sig$. We also show that, as in the Abelian
gaussian model, the averages of the connected current correlation
functions are unaffected by the disorder \cite{Be}. 

We are thus interested in computing the quenched average of the 
correlation functions of the currents $J^a_\mu= (\bar \Psi
\gamma_\mu t^a \Psi)$. Explicitely, their components are~:
\debut
J^a_z &=& \psi_{-;j} (t^a)^j_k \psi^k_+ \label{current}\\
J^a_\zb &=& \bar \psi_{-;j}  (t^a)^j_k \bar \psi^k_+ \non
\fin
While these currents are conserved in the pure system,
they are not conserved  once the disorder has been turned on. 
However, as explained in ref.\cite{Be}, the quenched theory still
possesses a $su(N)$ symmetry. The currents generating this
symmetry are represented by the insertion of the 
following operators $\CJ_\mu^a$ in the quenched averages~:
\debut
\CJ^a_\mu = \pi \frac{\de}{\de A^a_\mu} \non
\fin
Although the fields $J_z^a$ are not conserved, their
quenched correlation functions can be computed.
This relies on the Polyakov-Wiegman (PW) formula \cite{PW}
for the effective action $W[A]$ defined in eq.(\ref{partnona}).
It can be exactly computed by integrating 
its anomalous transformation under a chiral gauge transformation.
Indeed, let $G_\mu=\sum_at^aG_\mu^a[A]$ with
$G_\mu^a[A]=\vev{J_\mu^a}_A = \pi\frac{\de W[A]}{\de A^a_\mu}$ 
be the generating functions of the 
connected current Green function in the pure system.
They satisfy the anomalous Ward identities \cite{PW},
\debut
\d_z G_\zb + \d_\zb G_z +[A_z,G_\zb]+[A_\zb,G_z] &=& 0, \non\\
\d_z G_\zb - \d_\zb G_z +[A_z,G_\zb]-[A_\zb,G_z] &=&
2 F_{z\zb}[A] , \label{ano}
\fin 
with $F_{z\zb}[A]=\d_z A_\zb - \d_\zb A_z +[A_z,A_\zb]$.
Eqs.(\ref{ano}) completely specify $G_\mu[A]$. The solution of
eqs.(\ref{ano}) can written as~:
\debut
 G_z[A] &=& A_z - \inv{\d_\zb + ad.A_\zb} \d_zA_\zb \label{ga}\\
 G_\zb [A] &=& A_\zb - \inv{\d_z + ad.A_z} \d_\zb A_z \label{gza}
\fin
Here $ad.A_z$ denoted the adjoint action of $A_z$. Notice that
$G_z[A]$ is  local in $A_z$ but non-local in $A_\zb$.

Knowing explicitely the correlation functions for any impurity sample,
it is a priori possible to take the quenched average. Let
us first concentrate on the average of the chiral correlation
functions involving only currents of the same chirality; e.g.
involving only $J^a_z$. Consider first the average of products of
one-point functions. Since the quenched average
is defined by ${\bar {A_z^a(x)A_\zb^b(y)}}=\sig\pi\de^{ab}\de^{(2)}(x-y)$
and since $\vev{J_z^a}_A$ are linear in $A_z$, these quenched
correlations can be computed using Wick's theorem applied on $A$.
For example, the two-point and three-point functions are~:
\debut
{\bar {\vev{J_z^a(z_1)}_A\vev{J_z^b(z_2)}_A}}
&=& 2\sig\pi~\de^{ab}~ \({\inv{\d_\zb}~\d_z}\)_{(z_1,z_2)}
= \frac{ 2\sig \de^{ab} }{ (z_1-z_2)^2} \non\\
{\bar {\vev{J_z^a(z_1)}_A\vev{J_z^b(z_2)}_A\vev{J_z^c(z_3)}_A}}
&=& i(\pi\sig)^2 f^{abc} \({\inv{\d_\zb}~\d_z}\)_{(z_1,z_2)}
\[{\({\inv{\d_\zb}}\)_{(z_3,z_1)}-\({\inv{\d_\zb}}\)_{(z_3,z_2)}}\]\non\\
 && ~~~~ + (cyclic~~ permutation) \non\\
&=& 3\sig^2 \frac{if^{abc}}{(z_1-z_2)(z_1-z_3)(z_2-z_3)} \non
\fin

More generally, the average of products of one-point functions 
is the sum of connected correlations which can be expressed
in terms of the correlation functions of the pure system~:
\debut
\[{ {\bar {\vev{J_z^{a_1}(z_1)}_A \cdots \vev{J_z^{a_M}(z_M)}_A }} 
}\]^{connected} = M \sig^{M-1} \vev{J_z^{a_1}(z_1) \cdots J_z^{a_M}(z_M)}_0
\label{avone}
\fin
Here, $\vev{\cdots}_0$ denote the pure correlation functions.
They are known exactly \cite{KZ}. The relation between the quenched
correlations and their connected parts is the usual one. 

More interesting is the average of products of correlations
with insertion of the conserved currents $\CJ^a_\mu$ since they
encode the underlying symmetry algebra. We find~:
\debut
 &&{\bar {\CJ_z^{n_1}(z_1) \cdots \CJ_z^{n_M}(z_M)
\vev{J_z^{a_1}(w_1)J_z^{b_1}(\xi_1)\cdots}^c_A\cdots 
\vev{J_z^{a_P}(w_P)J_z^{b_P}(\xi_P)\cdots}^c_A }} \non\\
&=& \vev{J_z^{n_1}(z_1)\cdots J_z^{n_M}(z_M) 
J_z^{a_1}(w_1)J_z^{b_1}(\xi_1)\cdots}_0\cdots
\vev{J_z^{a_P}(w_P)J_z^{b_P}(\xi_P)\cdots}_0  \non\\
&+& \sum_{j=1}^M
\vev{J_z^{n_1}(z_1)\cdots {\widehat {J_z^{n_j}(z_j)}} 
\cdots  J_z^{n_M}(z_M) 
J_z^{a_1}(w_1)J_z^{b_1}(\xi_1)\cdots}_0\times\label{bigcor}\\
&& ~~~~~~\times
\vev{J_z^{n_j}(z_j) J_z^{a_2}(w_2)J_z^{b_2}(\xi_2)\cdots}_0 
\cdots \vev{J_z^{a_P}(w_P)J_z^{b_P}(\xi_P)\cdots}_0  \non\\
&+&\cdots  \non\\
&+& \vev{J_z^{a_1}(w_1)J_z^{b_1}(\xi_1)\cdots}_0\cdots
\vev{J_z^{n_1}(z_1)\cdots J_z^{n_M}(z_M)
J_z^{a_P}(w_P)J_z^{b_P}(\xi_P)\cdots}_0 \non
\fin
The hatted fields have to be omitted. 
Here, we assumed that there is no insertion of one-point functions.
The formula (\ref{bigcor}) is actually simpler in words~:
it is obtained by distributing the currents $J_z^{n_1}(z_1),
\cdots,J_z^{n_M}(z_M)$ among the pure correlators in all
possible way, each counted only once.
Notice that all the chiral quenched correlation functions are
purely algebraic, without any logarithmic correction.

Eq.(\ref{bigcor}) with no insertion of $\CJ^n_z$ shows that the
connected correlation function of the fields $J^a_z$ are
unaffected by the disorder.

 From eq.(\ref{bigcor}) we read the operator product expansion
of the fields. The currents $\CJ^a_z$ satisfy~:
\debut
\CJ_z^{n_1}(z_1) \CJ_z^{n_2}(z_2)
= \frac{\de^{n_1n_2}}{(z_1-z_2)^2}
+ \frac{if^{n_1n_2n_3}}{z_1-z_2} \CJ^{n_3}(z_2) + reg. \label{ope1}
\fin
Therefore, the quenched conserved currents satisfy the 
commutation relations of a Kac-Moody algebra, exactly
as the currents in the pure system do.

The operator product expansion between the conserved currents $\CJ^z_a$
and the correlators $\vev{J_z^{a_1}(w_1)J_z^{a_2}(w_2)\cdots}_A$ are~:
\debut
\CJ_z^{n}(z) \vev{J_z^{a_1}(w_1)J_z^{a_2}(w_2)\cdots}^c_A
&\sim& \sum_j \frac{\de^{na_j}}{(z-w_j)^2}
\vev{J_z^{a_1}(w_1)\cdots {\widehat {J_z^{a_j}(w_j)}}\cdots}^c_A \non\\
&~&~~+ \sum_j \frac{if^{na_ja'_j}}{z-w_j}
 \vev{J_z^{a_1}(w_1)\cdots J_z^{a'_j}(w_j)\cdots}^c_A
 + reg.\label{ope2}
\fin
These operator product expansions are unusual in conformal field
theory with Kac-Moody symmetry. In particular, they imply
that the fields  $\vev{J_z^{a_1}(w_1)J_z^{a_2}(w_2)\cdots}^c_A$
are {\it not} associated to highest weight vector representations.
We don't know to which category of representations they 
correspond to.

Contrary to the chiral quenched correlation functions which are easy to
compute, the averages of correlation functions involving fields
of opposite chiralities are difficult, if not impossible, to evaluate directly.
This is due to the fact that the generating functions $G_z[A]$
and $G_\zb[A]$ are non-local in $A_\zb$ and $A_z$ respectively.
A naive perturbative expansion is spoilled by untractable divergences.
However, since the model is asymptotically free in the
infrared, one can used renormalization group techniques
to evaluate few two-point functions \cite{Be}.

\bigskip
$\bullet$) {\bf Generalization to any WZW model.}

The previous discussion generalizes easily to any WZW models.
So let us consider a WZW model over the Lie group $\CG$ at level $k$.
Its action is \cite{Wit,KZ}~:
\debut
S_{wzw}(g) = k \int \frac{d^2x}{8\pi} tr(\d_z g^{-1}\d_\zb g) + k\Ga(g)
\non
\fin
where $\Ga(g)$ is the WZW topological term.
We define the action of the disordered WZW model to be the
WZW action in presence of sources. Namely~:
\debut
S_{wzw}(g;A) =  S_{wzw}(g) - k \int \frac{d^2x}{2\pi}
\({ tr(A_zJ_\zb+A_\zb J_z) + tr(gA_zg^{-1}A_\zb) - tr(A_z A_\zb)}\)\cr
\label{deswzw}
\fin
where $J_z, J_\zb$ are the WZW currents and $A_z,A_\zb$ the random variables.
We assume that the probability distribution $P[A]$ is given 
by eq.(\ref{mesnona}).

As a functional of $g$ and $A$ the action (\ref{deswzw}) satisfies the
``Polyakov-Wiegman" relation~:
\debut
S_{wzw}(g;A)= S_{wzw}(g^h;A^h) - S_{wzw}(hh^*;A^h),\quad
with ~~ \matrix{ A^h_\zb = h A_\zb h^{-1} + h\d_\zb h^{-1} \cr
                ~ \cr g^h = h g h^* \cr}
\label{gaugewzw}
\fin
for any element $h$ of the complexified group $\CG^C$.

If we parametrize the impurity connexion as in previous section
by $A_\zb= G^{-1}\d_\zb G$ with $G\in \CG^C$,
this action action has been cooked up such that~:
\debut
S_{wzw}(g;A)= S_{wzw}(g^G) - S_{wzw}(GG^*) 
\quad with \quad A_\zb= G^{-1}\d_\zb G=-A^*_z \label{wzwact}
\fin
This follows from eq.(\ref{gaugewzw}) and the fact that $A^G=0$.

This has two consequences. First we can explicitely
evaluate the partition function for any sample of the
disorder. Namely, we have~:
\debut
Z[A] = \int Dg e^{-S_{wzw}(g;A)}=\exp( S_{wzw}(GG^*)) \label{partwzw}
\fin
In the last equation we used the fact that for
the Haar measure on $\CG$ we have~:
$\int dg f(g_1gg_2^*) = \int dg f(g)$  for
any $g_1,~g_2\in\CG^C$ and any function regular enough.

The second consequence concerns the correlation
functions of the $\CG$-scalar local operators at fixed disorder.
For example consider the operator 
\debut
\Phi_R(z,\zb)=\sum_i \bar \phi_{\bar R; i}(\zb) \phi_R^i(z) \non
\fin
where $\phi_R^i(z)$ the primary fields in the representation $R$. 
These operators are invariant under the global $\CG$ symmetry.
In the path integral formulation, they are represented by
the insertion of $tr(\rho_R(g))$, ie~:
\debut
\Phi_R(z,\zb) \equiv tr(\rho_R(g)) \label{primwzw}
\fin
where $\rho_R$ denotes the representation $R$ of $\CG$. We then have~:
\debut
\vev{ tr(\rho_R(g)\cdots}_A &=& \int \frac{Dg}{Z[A]}~ e^{-S_{wzw}(g;A)}~
tr(\rho_R(g)) \cdots \cr \non
&=&  \int \frac{Dg}{Z[A]}~ e^{-S_{wzw}(g^G)+ S_{wzw(GG^*)}}~
tr(\rho_R(g)) \cdots \cr \non
\fin
where we have used eq.(\ref{wzwact}). Now we see from eq.(\ref{partwzw})
that the partition function $Z[A]$ cancels against $\exp(S_{wzw}(GG^*))$.
Therefore, using the invariance of the Haar measure, we obtain~:
\debut
\vev{ tr(\rho_R(g))\cdots}_A = \int Dg~ e^{-S_{wzw}(g)}~
tr(\rho_R(g)\rho_R(GG^*)^{-1}) \cdots \label{corwzw}
\fin
We have used the cyclicity of the trace in order to
reconstitute the product $GG^*$. In other words, we have used
the fact that the operator $tr(\rho_R(g))$ is $\CG$ invariant.
This is important because it proves that the $G$-dependence
of these correlation functions projects from
$\CG^C$ to $\CG^C/\CG$. Indeed, let us factorized $G\in \CG^C$ as~:
\debut
G= H~U \quad with\quad H\in \CG^C/\CG \quad and \quad U \in \CG
\label{gcgwzw}
\fin
Then since $UU^*=1$, we have~:
\debut
\vev{ tr(\rho_R(g)\cdots}_A = \int Dg~ e^{-S_{wzw}(g)}
tr(\rho_R(g)\rho_R(HH^*)^{-1}) \cdots \label{factowzw}
\fin
We thus have proved that the correlation functions of the
neutral local operators factorize into the WZW correlation
functions times an explicitely known function of the
$\CG^C/\CG$ component of the impurity connexion.
This is analogue to the factorization of the previous section.

Using the factorization (\ref{corwzw}) we can now average over $A$. 
As before, it is convenient to change variable from $A$ to $G=HU$. The 
Jacobian is given in eq.(\ref{jacobbis}). In particular at
$\sig=\infty$, we obtain~:
\debut
\bar { \vev{ tr(\rho_R(g)\cdots }_A\cdots}
= \int DH Dg~ e^{2h^vS_{wzw}(HH^*) - S_{wzw}(g) }~ 
tr(\rho_R(g)\rho_R(HH^*)^{-1}) \cdots \label{corwzwbis}
\fin
Therefore, we have proved the following equivalence, valid in
the $\CG$-neutral sector~:
\debut
\bar {{WZW_\CG^{(k)}}_A} = \({WZW}\)^\CN\times 
\({WZW_{\CG^C/\CG}^{(\tilde k =-2h^v)} }\) \label{equiwzw}
\fin
where $\CN$ is the number of correlation functions we are averaging.
These are conformal field theories. The conformal dimensions
of the operators in the disordered theory can be evaluate using the
formula (\ref{confcoset}).

\section{The supersymmetric method~:
  the random bond Ising model.}

In this section, we apply the supersymmetric method to study
the random bond Ising model and the closely related
Dirac theory coupled to a random potential and a random mass.
These models have been  analysed using the replica method in
ref.\cite{DoDo,Lu1,Shan}.

As we will soon explain, the randomness of the bond interaction
has a marginal effect on the large distance
behavior of the Ising model. 
This allows us to use
renormalization group techniques to study the infrared
behavior of this random model.  
At criticality, the disorder only induces 
logarithmic corrections to the pure system.
For example, the averages of the spin correlation functions
at criticality are \cite{Lu1,Shan}~:
\begin{eqnarray} 
{\bar {\vev{\sig(R) \sig(0)}^N} } \sim 
\(\frac{a}{R}\)^{N/4} \(\log(R/a)\)^{N(N-1)/8}\non
\end{eqnarray}
These have to be compared to the pure correlation
functin which are~: $\vev{\sig(R) \sig(0)}^N
\sim \(\frac{a}{R}\)^{N/4}$.

Close to criticality, the specific heat of the Ising
model possesses a logarithmic divergency~:
$C_v= \log(\tau)$ with $\tau\sim\frac{(T_C-T)}{T_C}$. In
the disordered model,
the behavior of the specific heat near criticality reads \cite{DoDo}~:
\debut
C_v \sim \inv{g_a} \log\( 1 +g_a \log(\inv{\tau})\) \non
\fin
where $g_a$ represents the strengh of the disorder.
This $\log(\log\tau)$ behavior has to be compared to the $(\log\tau)$
behavior in the pure system.

The perturbative study of the 
Dirac theory with a random potential and a random mass 
is very similar to the perturbative study
of the random mass Ising model.
However, the crucial difference is that contrary to 
the randomness of the mass, the randomness of the potential 
is marginally relevent.

Although the supersymmetric method is only applicable to
models which are gaussian at fixed disorder, it is certainly a
good starting point for disentangling properties of a large class
of disordered conformal field theories. 
The study of random bond Ising model clearly reveals that an appropriate
algebraic framework for studying gaussian disordered
systems at criticality should be based on affine Lie
superalgebra with zero superdimension. In the case of
the random Ising model, this algebra is $OSp(2N|2N)$.
The fact that the algebra has zero superdimension,
ie. has an equal number of bosonic and fermionic 
generators, ensures that the Virasoro central charge
of the Sugawara stress-tensor is zero. This is needed
by construction for a disordered system. The vanishing of the
central charge does not imply the triviality of the theory
since it is non-unitary. The general framework
should probably be based on the Wess-Zumino-Witten models, or their cosets,
on Lie superalgebras which, like $OSp(2N|2N)$, have equal numbers
of bosonic and fermionic generators.
The relevence of affine Lie superalgebras
was independently realized in ref.\cite{Murp}

\def\zb{ {\bar z} }
\def\psib{ {\bar \psi} }
\def\etab{ {\bar \eta } }
\def\gab{ {\bar \ga} }

\bigskip
$\bullet$) {\bf The model.}

In its scaling limit (near criticality) the Ising model is
described by a massive real Majorana fermion with mass
$\bar m\sim \tau = \frac{(T_C-T)}{T_C}$ with $T_C$ the critical temperature.
In presence of disorder, the mass becomes a function of the
space position.
The random Ising model is defined by the action $(z=x+iy)$~:
\begin{eqnarray} 
S[m(x)]=\int \frac{d^2x}{4\pi}
\({ \psi \d_\zb \psi + \psib \d_z \psib + im (x) \psib \psi }\)
\label{act1}
\end{eqnarray}
where $\psi$ are grassmanian variables.
The mass $m(x)$ is coupled to the energy operator~:
$\ep(x) = i\psib(x)\psi(x)$.
It is choosen to be a random quenched variable with a gaussian measure~:
\begin{eqnarray} 
P[m] = \exp\({- \inv{4g} \int \frac{d^2x}{2\pi}
(m(x)-\bar m)^2 }\)  \label{mesur}
\end{eqnarray}
The energy operator $\ep(x)$ has dimension one. The Harris criteria
thus tell us that randomness in the bond interaction is
marginal in the 2d Ising model. As we will see, it is not
exactly marginal but only marginally irrelevent.

Since, at fixed disorder the action (\ref{act1}) is a free
quadratic action,
we can use the supersymmetric formulation to study the 
quenched model. In order to compute averages of product of
correlation functions, we need to introduce an arbitrary
number of copies of the fermions and of their supersymmetric
partners. For simplicity, let us restrict to the case of
two copies. The generalization to an arbitrary number
of copies will be
given later. Let us denote by $\psi_1$ and $\psi_2$ 
the two real Majorana fermions, and introduce the complex
fermions $\psi_\pm$ by:
\begin{eqnarray} 
\psi_\pm = \inv{\sqrt{2}} (\psi_1 \pm i \psi_2)
\end{eqnarray}
At fixed disorder, the fermionic action is:
\begin{eqnarray} 
S_F &=& S_1[m(x)] + S_2 [m(x)] \non\\
 &=& \int \frac{d^2x}{4\pi}
 \({ \psi_1 \d_\zb \psi_1 +  \psi_2 \d_\zb \psi_2
 + \psib_1 \d_z \psib_1 + \psib_2 \d_z \psib_2
 + im (x) (\psib_1 \psi_1 + \psib_2 \psi_2)}\) \non\\
 &=& \int \frac{d^2x}{2\pi}\({
 \psi_- \d_\zb \psi_+ + \psib_- \d_z \psib_+
 +i\frac{m (x)}{2} (\psib_- \psi_+ - \psi_- \psib_+ )}\)
 \label{act2}
\end{eqnarray}
By definition this action can be used to compute products of two
Ising correlation functions. Namely~:
\debut
\vev{A(z)\cdots}_{m(x)}\vev{B(w)\cdots}_{m(x)}
= \inv{Z[m(x)]} \int D\psi_1 D\psi_2~ A_1(z)\cdots B_2(w)\cdots
e^{-S_F} \non
\fin
where $\vev{\cdots}_{m(x)}$ refers to the Ising correlation
function at fixed desorder. 
$A_1(z)$ and $B_2(w)$ are the expressions of the Ising operators
$A(z)$, $B(w)$ in the first and second copies.
The partition function is a determinant~:
\begin{eqnarray} 
Z[m(x)]=\int D\psi_\pm ~~e^{-S_F} 
= Det \pmatrix { -i\frac{m}{2} & \d_\zb \cr \d_z & i\frac{m}{2} \cr }
 = Det~H_{Dirac}\label{Dirac}
\end{eqnarray}
Its inverse can be represented as a path integral
over bosonic complex fields $\eta$ and $\ga$~:
\begin{eqnarray} 
\inv{Z[m(x)]} = \int D\eta D\ga ~~e^{-S_B} \non
\end{eqnarray}
with
\begin{eqnarray} 
S_B= \int \frac{d^2x}{2\pi}
\({ \eta\d_\zb \ga + \etab\d_z\gab + i\frac{m(x)}{2}(\etab\ga-\eta\gab) }\)
\label{act3}
\end{eqnarray}
Since the Dirac Hamiltonian $H_D$, as defined in (\ref{Dirac}),
is purely imaginary, $H_D^{\dag}=-H_D$,
the integral over the bosonic variables
is an integral of an imaginary gaussian if we defined the
complex conjugasion by $\eta^*=\gab~, \ga^*=\etab$.
To insure absolute convergence, we could add the term
$ \ep~\int \frac{d^2x}{\pi}(\etab\ga+\eta\gab)$
in the action $S_B$.

The total action is $S_{tot} = S_F + S_B $.
By construction, products of Ising correlation functions
at fixed disorder can be rewritten as~:
\debut
\vev{A(z)\cdots}_{m(x)}\vev{B(w)\cdots}_{m(x)} 
= \int D\psi_\pm D\eta D\ga~ A_1(z)\cdots B_2(w)\cdots
e^{-S_{tot}} \non 
\fin

The action $S_{tot}$ is supersymmetric. 
The supersymmetric transformation can be defined as~:
\debut
\delta \psi_- = \eta \quad&,&\quad \delta \psi_+=0\non\\
\delta \eta = 0 \quad&,&\quad \delta \ga =-\psi_+\non
\fin
Similarly for the $\bar \psi_\pm,\cdots$.
It is easy to check that the kinetic term in $S_{tot}$ is susy
invariant: $\delta(\psi_-\d_\zb\psi_+ + \eta\d_\zb \ga)=0$.
Moreover the field coupled to the mass terms in $S_{tot}$ is susy exact:
\debut
\({ \psib_- \psi_+ -\psi_- \psib_+  +\etab\ga-\eta\gab }\)
= \delta\( \psib_-\ga - \psi_-\gab\)\non
\fin
This insures that the partition function defined by $S_{tot}$
is independent of the random mass $m(x)$, exactly as BRS symmetry
ensures gauge invariance in gauge theories. We actually have
a stronger result~: $S_{tot}$ is susy-exact. 
This is similar to topological theory;
however, the difference with topological theory is that we are
interested in correlation functions of non susy-closed operators.
As we will explain below, 
this choice of the supersymmetric transformation is  not unique.

\bigskip
$\bullet$) {\bf The effective action.}

We now set $\bar m=0$. I.e. we consider the effect of the
disorder at criticality.
Averaging over the quenched variables with the
gaussian measure (\ref{mesur}) leads to the effective action~:
\begin{eqnarray} 
S_{eff} &=& \int \frac{d^2x}{2\pi}\({
\psi_- \d_\zb \psi_+ + \psib_- \d_z \psib_+
+\eta\d_\zb \ga + \etab\d_z\gab }\)
+\frac{g}{8} \int\frac{d^2x}{\pi}~\Phi_{pert} \non\\
&=& S_* + \frac{g}{8} \int\frac{d^2x}{\pi}~\Phi_{pert} 
\label{Seff1}
\end{eqnarray}
with
\begin{eqnarray} 
\Phi_{pert}&=&\({ \psib_- \psi_+ - \psi_- \psib_+ 
+\etab\ga-\eta\gab }\)^2 \label{pert1}\\
&=& 2 (\psib_1\psi_1)(\psib_2\psi_2) + (susy~partners).\non
\end{eqnarray}
Clearly this action is supersymmetric since the
total action $S_{tot}$ was supersymmetric for any values of the
disorder before averaging.

This action can be viewed as a perturbation of the (non-unitary)
conformal field theory specified by the action $S_*$.
This fixes the normalization of the fields to be~:
\begin{eqnarray} 
\psi_-(z) \psi_+(w) \sim \inv{z-w} \quad,\quad
\ga(z)\eta(w) \sim \inv{z-w} \label{norm}
\end{eqnarray}
The Virasoro algebra is the standard one. The Virasoro
central charge is zero. Note that since
the fermions $\psi_\pm$ have dimension half, the perturbing
field  $\Phi_{pert}$ has dimension two.
It is therefore marginal.

The conformal field theory specified by $S_*$ is
invariant under a supersymmetric algebra whose
conserved currents are~:
\debut
G_\pm(z) = \eta(z)\psi_\pm(z) \quad&,&\quad
\hat G_\pm(z) = \ga(z)\psi_\pm(z) \non\\
K(z) = \eta^2(z) \quad&,&\quad \hat K(z)=\ga^2(z) \non\\
J(z)=:\psi_-(z)\psi_+(z): \quad&,&\quad
H(z)=:\ga(z)\eta(z):\label{curr}
\fin
There are four fermionic currents, which are generators
of supersymmetric transformations, and four bosonic ones.
They form a representation of the affine $OSp(2|2)$
current algebra at level one. The root system of
$OSp(2|2)$ is given below~:
\debut
\matrix{ ~~&~~~&\otimes_{G_+}&~~&~~&~~&\otimes_{\hat G_+}&~~&~~\cr
         ~~&~~~&~~&~~&~~&~~&~~&~~&~~\cr
	\bullet_K&~~&~~&~~&\bullet_J\bullet_H&~~&~~&~~&\bullet_{\hat K}\cr
        ~~&~~~&~~&~~&~~&~~&~~&~~&~~\cr
        ~~&~~~&\otimes_{G_-}&~~&~~&~~&\otimes_{\hat G_-}&~~&~~\cr}\non
\fin
The black dots refer to the bosonic generators, while the symbols
$\otimes$ represent the fermionic roots.
The non-trivial operator product expansions of the currents
are the following~:
\debut
J(z)J(w) \sim \frac{1}{(z-w)^2} 
\quad &;& \quad
H(z)H(w) \sim \frac{-1}{(z-w)^2} \non\\
J(z)G_\pm(w) \sim \frac{\pm1}{(z-w)} G_\pm(w)
\quad &;& \quad
J(z)\hat G_\pm(w) \sim \frac{\pm1}{(z-w)} \hat G_\pm(w)\non\\
H(z)G_\pm(w) \sim \frac{1}{(z-w)} G_\pm(w)
\quad &;& \quad
H(z)\hat G_\pm(w) \sim \frac{-1}{(z-w)} \hat G_\pm(w)\label{opecurrent}\\
H(z) K(w) \sim \frac{2}{(z-w)} K(w)
\quad &;& \quad
H(z)\hat K(w) \sim \frac{-2}{(z-w)}\hat K(w)\non\\
\hat G_\pm(z) G_\mp(w) &\sim& \frac{1}{(z-w)^2}
+ \frac{1}{(z-w)}(H(w)\pm J(w)) \non\\
\hat K(z) K(w) &\sim& \frac{2}{(z-w)^2}
+\frac{4}{(z-w)} H(w) \non\\
G_-(z) G_+(w) \sim \frac{1}{(z-w)} K(w)
\quad &;& \quad
\hat G_-(z) \hat G_+(w) \sim \frac{1}{(z-w)} \hat K(w) \non\\
K(z) \hat G_\pm(w) \sim \frac{-2}{(z-w)} G_\pm(w)
\quad &;& \quad
\hat K(z) G_\pm(w) \sim \frac{2}{(z-w)} \hat G_\pm(w)\non
\fin

The perturbing field can be written in terms of
these fields~:
\debut
\Phi_{pert}=2\[{ \bar J J - \bar H H +\half(\bar K \hat K
+\bar {\hat K} K) + \bar G_- \hat G_+ -
\bar {\hat G_-} G_+ + \bar G_+ \hat G_- - \bar {\hat G_+} G_-}\]\non
\fin
In other words, the perturbation (\ref{pert1}) is
a current-current perturbation. It preserves the
$OSp(2|2)$ symmetry.

The generalization to $2N$ copies is obvious. We introduce $2N$ copies of
Majorana fermions $\psi_\al$, $\al=1,\cdots,2N$, or $N$ copies of
Dirac fermions $\psi_\pm^\al$, $a=1,\cdots,N$,
and $2N$ copies of their supersymmetric partners.
The perturbing field is similarly obtained by a gaussian
integral from the free random theory~:
\begin{eqnarray} 
\Phi_{pert} =
2\sum_{\al<\beta} (\psib_\al\psi_\al) (\psib_\beta\psi_\beta) 
+ (susy~partners) \label{pert2}
\end{eqnarray}
The symmetry algebra is now extended to the affine $OSp(2N|2N)$
current algebra. The perturbing field is still bilinear in
the currents. Hence, the random Ising model 
is described by a Gross-Neveu model on the Lie superalgebra
$OSp(2N|2N)$. In particular  it is an integrable model,
with factorizable scattering,  Yangian symmetry, etc ...

\bigskip
$\bullet$) {\bf Renormalization group computation.}

The action $S_{eff}$ describes an interacting theory.
Although the perturbing field has dimension two,
and therefore is naively marginal,
it breaks conformal invariance. To see it requires
computing the beta function. As explained in
the Appendix, the beta function at one loop is encoded in
the operator product expansion (OPE) of the perturbing field.
Since $S_*$ is a free gaussian theory, this OPE is easily
computed using  Wick's theorem and  eqs. (\ref{norm}).
We obtain~:
\debut
\Phi_{pert}(z) \Phi_{pert}(w) = \frac{8}{|z-w|^2}~\Phi_{pert}(w)
+ {\rm irrelevent ~ terms.} 
\non
\fin
Using the formula (\ref{A3}) of the Appendix, we get the beta function~:
\begin{eqnarray} 
\dot g =\beta(g) = -g^2 + \cdots \label{beta1}
\end{eqnarray}
The dots refer to higher loop corrections.
We recognize the beta function for asymptotically free theory.
It is easily integrated, giving the coupling constant flow~:
\begin{eqnarray} 
g_R = \frac{g_a}{1 + g_a\log(R/a)} \non
\end{eqnarray}
where $g_a$ is the value of the coupling
constant at the lattice cut-off $a$. Since $g_a$ is
positive by definition (it is the strengh of the disorder),
$g_R$ decreases at large distances $R$. In other words, the
theory is asymptotically free in the infrared regime.

The beta function is independent of the number of copies, ie. of $N$.

\bigskip
$\bullet$) {\bf Quenched correlation functions.}

The fact that the random Ising model is 
asymptotically free in the infrared regime is important
since it allows us to compute the large distance behavior
of two point quenched correlation functions using the renormalization
group, (using formula (\ref{A4}) of the Appendix
for the anomalous dimensions).

We recall the OPE between the energy operator
and the spin field. The fusion rules are
$\ep\times \sig=\sig$. More precisely, we have~:
\begin{eqnarray} 
\ep(z)\sig(w)=
i(\psib\psi)(z) \sig(w)= \frac{C_{\ep\sig\sig}}{|z-w|} \sig(w)
+ \cdots \non
\end{eqnarray}
with $C_{\ep\sig\sig}^2=1/4$.

Let us consider the quenched average of products of 
two point functions of the spin field, Ie.
${\bar {\vev{\sig(R) \sig(0)}^N} }$. Consider first the case $N=1$. 
In the supersymmetric effective theory, the quenched
correlation function ${\bar {\vev{\sig(R) \sig(0)}} }$
is represented by the two point function of the
spin  field $\sig_1$ in the first copy. Its
OPE with the perturbing field is ~:
\begin{eqnarray} 
\Phi_{pert}(z) \sig_1(w) = \inv{|z-w|} \CO(w) + \cdots
\non
\end{eqnarray}
where $\CO(w)$, an operator of dimension $(1/8+1)$,
is irrelevent compared to $\sig_1$ which
has dimension $1/8$.
Therefore, the anomalous dimension of $\sig_1$ at one
loop is $\ga^{(1)}= 1/8 + \cdots$, up to irrelevent
terms. Hence:
\begin{eqnarray} 
{\bar {\vev{\sig(R) \sig(0)}} }
\sim \({\inv{R}}\)^{1/4} \label{coor1}
\end{eqnarray}
It has the same infrared behavior as the spin-spin
correlation function in the pure system at the critical
temperature.

Consider now the general case $N$ arbitrary.
In the supersymmetric effective action,
the quenched correlation function ${\bar {\vev{\sig(R) \sig(0)}^N} }$
is represented by the two point function of the operator
$\CO^{(N)}(z)$, product of the spin fields in the $N$
first copies~:
\begin{eqnarray} 
\CO^{(N)}(z) = \sig_1(z)\cdots \sig_N(z) \non
\end{eqnarray}
When computing the OPE between the perturbing field $\Phi_{pert}$
and $\CO^{(N)}(z)$ only the first term in eq.(\ref{pert2}),
$2\sum_{\al<\beta} (\psib_\al\psi_\al) (\psib_\beta\psi_\beta)$,
gives a relevent contribution. 
When computing the OPE between this term and $\CO^{(N)}(z)$
we can contract any pair of products of 
$(\psib_\al\psi_\al) (\psib_\beta\psi_\beta)$
with the pair of product of spin fields $\sig_\al\sig_\beta$.
E.g.
\debut
(\psib_1\psi_1)(\psib_2\psi_2)~ \sig_1(z)\sig_2(z)\cdots \sig_N(z)
\non
\fin
Thus we have~:
\debut
\Phi_{pert}(z) \CO^{(N)}(w)=
-\frac{N(N-1)C_{\ep\sig\sig}^2}{|z-w|^2}  \CO^{(N)}(w)
+ irrelevent~terms. \non
\end{eqnarray}
Therefore, the anomalous dimension of $ \CO^{(N)}(w)$ is
\begin{eqnarray} 
\ga^{(N)}= \frac{N}{8} - \frac{N(N-1)C_{\ep\sig\sig}^2}{4}~g +\cdots\non
\end{eqnarray}
Recall that $C_{\ep\sig\sig}^2=1/4$.
It implies~:
\begin{eqnarray} 
{\bar {\vev{\sig(R) \sig(0)}^N} }_{g_a}&=&
\exp\({ -2\int^{\sig_R}_{g_a} \frac{\ga^{(N)}}{\beta} d\sig}\)~
{\bar {\vev{\sig(a) \sig(0)}^N} }_{g_R} \non\\
&\sim& \(\frac{a}{R}\)^{N/4} \(\log(R/a)\)^{N(N-1)/8}
\label{corr2}
\end{eqnarray}
There are logarithmic corrections to the pure system.

\bigskip
$\bullet$) {\bf The specific heat near criticality.}

In order to compute the behavior of the mean specific heat near
the critical point we consider a non zero value of $\bar m$:
\debut
\bar m = \frac{ T_C - T}{T_C} = \tau\non
\fin
The specific heat $C_v$ per unity of volume 
$C_v= \inv{Vol.}\({\frac{\d E}{\d T}}\)$ can be expressed in terms
of the connected correlation function of the energy operator $\ep(x)$~:
\debut
C_v = \int d^2x~ {\bar {\vev{\ep(x)\ep(0)}^{conn.}}}\non
\fin
Near criticality, its behavior is given by the infrared singular
behavior of ${\bar {\vev{\ep(x)\ep(0)}^{conn.}}}$ at zero mass.
The infrared cut-off is specified by the mass scale $1/\bar m$.
Thus~:
\debut
C_v \sim \int^{1/\bar m} d^2x~~ 
{\bar {\vev{\ep(x)\ep(0)}^{conn.}}_{\bar m=0}}\label{heat1}
\fin

In the supersymmetric effective theory, the connected two-point 
correlation function for the energy operator is represented by the
two-point function of the operator $\ep'(x)= \ep_1(x)-\ep_2(x)$
where $\ep_{1,2}(x)$ refer to the energy operators in the first
and second copies~:
\debut
2 {\bar {\vev{\ep(x)\ep(0)}^{conn.}}} = \vev{\ep'(x)\ep'(0)}\non
\fin
To evaluate the behavior of $C_v$ as $\bar m\to 0$ requires
evaluating the infrared behavior of $ \vev{\ep'(x)\ep'(0)}$.
This can be done using the renormalization group by 
computing the anomalous dimension of $\ep'(x)$.
The OPE between $\ep'(x)$ and the perturbing field $\Phi_{pert}$
follows from the fusion rule, $\ep\times\ep = 1+ \ep$, in
the Ising model~:
\debut
\Phi_{pert}(z) \ep'(w) = \frac{2}{|z-w|^2} \ep'(w) + \cdots \non
\fin
Formula (\ref{A4}) of the Appendix gives~:
\debut
\ga_{\ep'}(g) = 1 + \half g + \cdots \non
\fin
Therefore, at large distances we have~:
\debut
\vev{\ep'(R)\ep'(0)} \sim \(\frac{a}{R}\)^2 
\inv{(1+g_a\log(R/a))} \non
\fin
This gives the critical behavior of the specific heat~:
\debut
C_v &\sim& \int^{1/\tau} d^2x
{\bar {\vev{\ep(x)\ep(0)}^{conn.}}}\non\\
&\sim& \inv{g_a} \log\( 1 +g_a \log(\inv{\tau})\) \label{heat2}
\fin
This $\log(\log\tau)$ behavior has to be compared to the $(\log\tau)$
behavior in the pure system.

\bigskip
$\bullet)$ {\bf Miscellaneous remarks on the random Dirac theory
\footnote[4]{These remarks arised from discussions we had with
Martin Zirnbauer.}.}

A model very closely connected to the random
bond Ising model has been introduced in connection with 
the quantum Hall transition \cite{Luetal}. It is a model of
Dirac fermions coupled to a random potential, or more
generally, to a random vector potential, a random mass and
a random scalar potential. Its action is~:
\debut
 S &=&\int \frac{d^2x}{2\pi}\({
  \psi_- (\d_\zb + A_\zb)\psi_+ + \psib_- (\d_z+A_z) \psib_+ }\right.\non\\
 &&\left.{+i\frac{m(x)}{2} (\psib_- \psi_+ -\psi_- \psib_+ )
 +i\frac{V (x)}{2} (\psib_- \psi_+ +\psi_- \psib_+ )}\)
 \non
\end{eqnarray}
The random variables $A$, $m$ and $V$ have a
gaussian distribution with width $g_A$, $g_M$ and $g_V$.
Let us denote by $\Phi_A$, $\Phi_M$ and $\Phi_V$ the
perturbing fields coupled to the constants
$g_A$, $g_M$ and $g_V$ after averaging over the disorder.
In the two copy sector, we can write them in terms of
the $OSp(2|2)$ currents as follows~:
\debut
\Phi_A&=& (\bar H-\bar J)(H-J) \non\\
\half(\Phi_V+\Phi_M) &=& 2 \bar G_- \hat G_+ - 2 \bar {\hat G_+} G_-
 + \bar K \hat K + \bar {\hat K} K \non\\
\half(\Phi_V-\Phi_M) &=& 2 \bar H H - 2 \bar J J + 2\bar {\hat G_-} G_+
 -2 \bar G_+ \hat G_- \non
\fin

It is an interesting exercise to compute the beta
function at one-loop for these coupling constants. We find~:
\debut
\beta_A= \dot g_A &=& 32~ g_M~g_V +\cdots\non\\
\beta_V-\beta_M=\dot g_V - \dot g_M &=& 8~(g_V+g_M)^2 + \cdots\label{flowdir}\\
\beta_V+\beta_M=\dot g_V + \dot g_M &=& 8 (g_V+g_M)(g_V-g_M) +
8 g_A(g_V+g_M) + \cdots \non
\fin
For generic value of the initial coupling constants this
describes  complicated flows. In the particular case
with $g_A=g_V=0$, we recover the beta function of the
random bond Ising model. 

For $g_A=g_M=0$, we have $\dot g_V= 8 g_V^2$. This 
means that  a random potential is marginally relevent in the
Dirac theory \cite{Luetal}. The coupling constant $g_V$ grows at large
distances. The infrared fixed is expected to describe the
quantum Hall effect transition. However, being  a strong
coupling problem no concrete description of it has
been proposed. 

Note that in this perturbative computation we put all the 
epsilons which distinguished between advanced and retarded
sectors equal to zero. In other words, we did not distinguished
whether we are computing averages of product of only
advanced Green functions, or averages of product of advanced
and retarded Green functions. It is likely that the 
infared behavior will not be the same in these two cases.
It is tempting to conjecture that it will be trivial 
when we consider only advanced Green functions but
non-trivial when mixed products of advanced and retarded
Green functions are considered.

Being formulated as current-current perturbation of a first 
order free theory, we can implement a Hubbard-Stratonovitch 
transformation on the random Dirac theory
 in order to produce a dual theory whose 
coupling constant is $1/g$. The transformation goes 
as follows~: one first represents the current-current interaction
by a gaussian integral over an auxiliary gauge connexion $A$, 
and then integrate over the free fermions and bosons. 
This produces a dual theory whose fundamental fields are
the connexion $A$, or equivalently the group element 
$G$ such that $A_\zb=G^{-1}\d_\zb G$, and whose  
fundamental coupling constant is $1/g$. 
However, in the mixed sector with both advanced and retarded 
Green functions, this naive transformation seems to be
spoiled by divergences in the gaussian integrals.

\section{The replica method~: 
the random phase sine-Gordon model.}

In this section, we apply the replica method to the 
random phase sine-Gordon model.
It is a random version of the 2d XY model.
It has also been used to describe other
random physical systems; e.g. the 2d XY model in
a random field \cite{CaOs,VilFer}, interfacial roughening transition \cite{CuSha},
randomly pinned flux lines in supraconductors \cite{Fish}, etc...

Let us first introduce the model from the XY point of view.
Recall that the partition function of the XY model
is defined as~:
\debut
Z= \sum_{\{h_i\}} \exp\({ - \frac{K}{2}
\sum_{\vev{i,j}}(h_i-h_j)^2 }\) \non
\fin
It describes the thermodynamics of a fluctuating
surface. The variables  $h_i$ are interpreted as
the height of the surface above a base plane.
In the pure system the $h_i$ take integer values.
The disordered model is defined by the same partition
function but the height variables $h_i$ take the values
$h_i=d_i+n_i$ with $n_i$ integers and $d_i$ some random
variables. Similarly as for the standard XY model,
we can formulate this problem in an alternative form
by using the Poisson summation formula~:
$\sum_n f(n)= \sum_m \int d\Phi f(\Phi)e^{i2\pi m\Phi}$.
Introducing the variables $\Phi_i$, one gets~:
\debut
Z= \int D\Phi \sum_{\{m_i\}} \exp\({ -\frac{K}{2}
\sum_{\vev{i,j}}(\Phi_i-\Phi_j)^2 + i2\pi
\sum_j m_j(\Phi_j-d_j)}\)\non
\fin
In the continuum limit, the variables $\Phi_i$ are replaced 
by a field $\Phi(x)$ and the discrete Laplacian becomes
a continuous one. Near the critical point, only the
first harmonic of the local potential is relevent. 
Keeping only this first harmonics gives the action~:
\debut
S =\int \frac{d^2x}{4\pi}\({ \frac{K}{2}
(\d_\nu\Phi)^2 - \Delta \cos(\Phi(x)-d(x))}\)\non 
\fin
with $d(x)$ a random quenched fields.
This is the model we will use to illustrate the replica method.

As we explain below, this model as two different phase~: a high
temperature phase in which the disorder is irrelevent and a
low temperature phase in which it is relevent. In the 
high temperature phase nothing interesting happens. 
In the low temperature phase, the large distance behavior of the 
system could a priori be different than in the pure system.
However, a precise description of this behavior is still
missing. For example, the behavior of the quenched correlation
functions of $\Phi(x)$
is still controversial. One finds that at large distance
we have~:
\debut
 \bar {\vev{~[\Phi(x)-\Phi(0)]^2~}} =
A(\log|x|) + B(T-T_c)^2 (\log|x|)^2\non
\fin
However, renormalization group computations predict a
non-vanihsing $B\not= 0$, while variational approaches
give $B=0$. In the sequel we will present both approaches.

\bigskip
$\bullet$) {\bf The model.}

Let us first rewrite the action with more appropriate notations.
We also need to slightly generalize it by introducting a
random potential in addition to the random phase.
The action reads~:
\debut
S(\Phi|A_\mu;\xi)
= \int \frac{d^2x}{4\pi}\({ \frac{K}{2} 
(\d_\nu\Phi)^2 - A_\nu \d_\nu\Phi - \xi(x) e^{i\Phi} 
-\xi^*(x) e^{-i\Phi} }\) \label{actA}
\fin
The constant $K$ is proportional to the inverse temperature, 
$K\propto 1/T$.
In the formula, $A_\nu(x)$ and $\xi(x)$ are random quenched variables with
gaussian measure~:
\debut
P[A] = \exp\[{ -\inv{2g} \int \frac{d^2x}{4\pi} A_\mu A_\mu}\]\non\\
P[\xi] = \exp\[{ -\inv{2\sig} \int \frac{d^2x}{4\pi}~\xi \xi^* }\] 
\label{gauss1}
\fin

In absence of disorder, the dimension of the vertex
operator $\exp(i\Phi)$ is~: $\dim( e^{i\Phi}) = \inv{K}$. 
The Harris criteria then
tell us that there are two different phases~:
\debut
low~temperature~phase~(K>K_c=1) &\Rightarrow& disorder~is~relevent\non\\
high~temperature~phase~(K<K_c=1) &\Rightarrow& disorder~is~irrelevent\non
\fin
Recall that $K\propto 1/T$.
At high temperature, the large distance behavior is identical to
that of the pure system.
At the critical temperature $K=K_c$, the disorder has a marginal effect.
At low temperature, the large distance behavior is a priori different.

\bigskip
$\bullet)$ {\bf Symmetries.}

Similarly as for the random gaussian model, the random phase
sine-Gordon model possesses a $U(1)$ symmetry whose
Noether current corresponds to insertion of
$\d_\mu\Phi$ in the quenched connected correlation functions.
This symmetry in particular implies that all the quenched
connected correlation functions of $\d_\mu\Phi$
are unaffected by the disorder.

 Just as for the Gaussian model, we can decompose $A_\mu$ as
$A_\mu=\d_\mu \La + \ep_{\mu\nu}\d_\nu\zeta$. Then the  field $\zeta$
decouples from the action and from the measure. So we can set it to zero,
and we have~:
\debut
P[A] = \exp\[{ -\inv{2g} \int \frac{d^2x}{4\pi}
(\d_\mu\La)^2 }\],\quad with\quad A_\mu=\d_\mu \La
\label{mesurla}
\fin
We will denote by $S(\Phi|\La;\xi)$ the action (\ref{actA}) with
$A_\mu=\d_\mu \La$.

The simplest way to visualize this symmetry consists in
introducing sources for $\d_\mu\Phi$ in the action (\ref{actA})~:
\debut
S(\Phi|\La;\xi) \to S(\Phi|J_\mu;\La;\xi)=
S(\Phi|\La;\xi) - \int \frac{d^2x}{4\pi} J_\mu\d_\mu\Phi \non
\fin
Let us decompose $J_\mu$ as $J_\mu=\d_\mu\rho + \ep_{\mu\nu}\d_\nu\eta$.
The field $\eta$ decouples and only the field $\rho$ is
relevent. The $U(1)$ symmetry follows form 
the fact that the $\rho$ and $\La$ dependence can be
absorbed into a shift of $\Phi$. Namely~:
\debut
S(\Phi|J_\mu;\La,\xi) = S(\Phi-\frac{\rho+\La}{K}|J_\mu=\La=0,\hat \xi)
-\inv{2K} \int \frac{d^2x}{4\pi} (\d_\mu(\rho+\La))^2 \label{shift}
\fin
with $\hat \xi= \xi e^{i(\rho+\La)/K}$ and $J_\mu=\d_\mu\rho$.
As a consequence, the generating function of the connected
correlation functions of $\d_\mu\Phi$, which is $\log Z[J_\mu,\La,\xi]$,
can be written as~:
\debut
\log Z[J_\mu,\La,\xi] = 
-\inv{2K} \int \frac{d^2x}{4\pi} (\d_\mu(\rho+\La))^2
+ \log Z[J_\mu=\La=0,\hat \xi] \label{symu1}
\fin
Now, notice that $\hat \xi$ and $\xi$ have identical distribution.
Therefore, averaging (\ref{symu1}) over the disorder implies~:
\debut
\bar {\log Z[J_\mu,\La,\xi]} = 
-\inv{2K} \int \frac{d^2x}{4\pi} (\d_\mu\rho)^2 + const. \non
\fin
where the constant is independent of $J_\mu$.
This shows that the quenched average of the connected correlation functions
of the $U(1)$ current are identical to the correlation functions
of this current in absence of disorder. In other words, the
connected correlation functions are protected from the disorder
by the $U(1)$ symmetry.

It is worth noticing that the presence of the disorder restore the
$U(1)$ symmetry which is absent in the pure sine-Gordon model.

In connection with this $U(1)$ symmetry, the model possesses
the remarkable property that
the $g$-dependence of the correlation functions
of the vertex operators can be factorized.
Consider eq.(\ref{shift})  in absence of source~: $J_\mu=0$.
The fact that the $\La$-dependence can be absorbed in a translation
of $\Phi$ implies for the correlation functions that~:
\debut
\vev{ e^{i\al_1\Phi(x_1)}\cdots}_{\La,\xi}
= \({\prod_p e^{i\frac{\al_p}{K}\La(x_p)}}\)~ 
\vev{ e^{i\al_1\Phi(x_1)}\cdots}_{\La=0,\hat \xi} \non
\fin
Let $G_{\al_1,\cdots}(x_1,\cdots|g,\sig)$ be the 
quenched correlations of the vertex operator $\exp(i\al\Phi)$~:
\debut
G_{\al_1,\cdots}(x_1,\cdots|g,\sig)
=\bar {\vev{ e^{i\al_1\Phi(x_1)}\cdots}_{\La,\xi}
\vev{\cdots}_{\La,\xi}} \label{vertex}
\fin
Integrating over $\La$ using the free field gaussian measure
(\ref{mesurla}), and using the fact that $\hat \xi$ and $\xi$
have the same measure, we deduce~:
\debut
G_{\al_1,\cdots}(x_1,\cdots|g,\sig)
=\prod_{p<q}|x_p-x_q|^{2g\al_p\al_q/K^2}~~
G_{\al_1,\cdots}(x_1,\cdots|g=0,\sig) \label{Gg}
\fin
Equivalently,
\debut
\d_g G_{\al_1,\cdots}(x_1,\cdots|g,\sig)
=\({\sum_{p<q} \frac{\al_p\al_q}{K^2} \log(|x_p-x_q|^2)}\)~
G_{\al_1,\cdots}(x_1,\cdots|g,\sig) \label{gdepend}
\fin
This identity will be useful for analyzing the
renormalization group equations.

\bigskip
$\bullet$) {\bf The effective action.}

Since the action (\ref{actA}) is not free we cannot rely on the
supersymmetric method but only on the replica trick. 
Therefore, as explained in the introduction, we consider $n$ copies 
of the system with the same disorder and then average over the
disorder. This gives the following effective action~:
\debut
S_{eff}= \int \frac{d^2x}{4\pi}\({
\frac{K}{2}\sum_r (\d_\mu\Phi^r)^2 - \frac{g}{2}
\sum_{r,s} (\d_\mu\Phi^r)(\d_\mu\Phi^s)
- 2\sig \sum_{r\not= s} \exp(i(\Phi^r-\Phi^s)) }\)
\label{actrsg}
\fin
Let us first look at the kinetic term. 
It is of the form $\int \frac{d^2x}{4\pi}\half
 (\d_\mu\Phi^r) G_{rs}(\d_\mu\Phi^s)$ with
\debut
G_{rs}= K\de_{rs}-g = (K-g)\de_{rs}-g(1-\de_{rs})
\non
\fin
Since the interaction only involves the difference of
the replicated fields, it is convenient to
decompose the kinetic terms as~:
\debut
\half  (\d_\mu\Phi^r) G_{rs}(\d_\mu\Phi^s)
= \frac{(K-ng)}{2n}\({\d_\mu (\sum_r\Phi^r)}\)^2
+ \frac{K}{4n} \sum_{r\not= s} \({\d_\mu(\Phi^r-\Phi^s)}\)^2
\label{kine}
\fin

Note that the field $\({\sum_r\Phi^r}\)$  decouples.
Its correlation functions are therefore unaffected
by the disorder. Since these correlation functions
represent the averages of the connected correlation
functions of the $U(1)$ current, we recover the previous
result obtained from the $U(1)$ Ward identities.
In particular, $\d_\mu (\sum_r\Phi^r)$ is a $U(1)$ conserved
current.
We could also use this decomposition to derive eq.(\ref{gdepend}).

The kinetic term (\ref{kine}) fixes the normalization of
the vertex operators. In particular, we find the
dimension of the perturbing field~:
$\dim\({e^{i(\Phi^r-\Phi^s)}}\) = \frac{2}{K}$. 
It is independent of $\sig$.  It is relevent for $K>1$~:
we thus recover the Harris criteria.

This decomposition has a Lie algebraic interpretation. It 
corresponds to the decomposition of $U(n)=SU(n)\times U(1)$.
The second term in (\ref{kine}) can be rewritten
as $\frac{K}{2} (\d_\mu\vec {\varphi})^2$ where
$\vec {\varphi}$ takes values in the Cartan subalgebra
of $SU(n)$. In terms of $\vec {\varphi}$, the perturbing
field in (\ref{actrsg}) reads~:
$\sum_{\vec \al} \exp(i\vec \al .\vec {\varphi})$
where the sum extends over all the $SU(n)$ roots.
At $K=K_c$, the fields $\exp(i\vec \al .\vec {\varphi})$
have dimension two, and the perturbing field is a current-current
interaction. The action (\ref{actrsg}) at $K=K_c$ is therefore equivalent
to the $SU(n)$ Gross-Neveu model. We will later see that at this
point the model is asymptotically free in the infrared.
Hence at the critical temperature $K=K_c$, the disorder
only induces logarithmic corrections.

\bigskip
$\bullet$) {\bf Renormalization group.}

The renormalization group allows to perturbatively
analyse the behavior of the system in the low temperature
phase. We will do a one loop computation, which is
valid close to the critical temperature,
ie. $\frac{K-K_c}{K_c}\ll 1$.

As we are already familiar with, the one-loop beta functions
are encoded in the OPE of the fields. Let us introduce
the following notation~:
\debut
\CO_1 &=& \sum_{r\not= s} \exp(i (\Phi^r-\Phi^s)) \label{O1}\\
\CO_2 &=& \inv{2n} \sum_{r,s} i\d_z(\Phi^r-\Phi^s)i\d_\zb(\Phi^r-\Phi^s)
\label{O2}
\fin
The field $\CO_1$ is the perturbing field associated to the
coupling constant $\sig$. The field $\CO_2$ is one of the kinetic field.
Notice that $\CO_2= (i\d_z\vec\varphi)(i\d_\zb\vec\varphi)$.
It is necessary to introduce it since
it is generated from $\CO_1$ by OPE. Indeed we have~:
\debut
\CO_1(z)\CO_1(0) &=& \frac{2(n-2)}{|z|^{2/K}} \CO_1(0)
+ \frac{ 2n}{|z|^{2/K-2}} \CO_2(0) 
+ irrelevent~ terms,\label{opesg}\\
\CO_2(z)\CO_1(0) &=& \frac{2/K^2}{|z|^2} \CO_1(0) 
+ irrelevent~ terms, \non\\
\CO_2(z)\CO_2(0) &=& irrelevent~ terms.\non
\fin
For $n=1$ these are the familiar OPE of the Kosterlitz-Thouless
transition. Note the change of behavior between $n<2$ or $n>2$.
The first OPE indicates how $\sig$ is renormalized, but the
second equation shows that the kinetic term needs also
to be renormalized.

Using the formula (\ref{A3}) of the Appendix, we get the beta
functions~:
\debut
\beta_\sig &=& 2\({\frac{K-K_c}{K}}\)~ \sig +(n-2)~\sig^2 + \cdots 
\non\\
\beta_K&=&  \frac{n}{2}~\sig^2 +\cdots \non
\fin
This shows that the coupling to $A_\mu$ would
have been generated at one loop even if we did not start with it.
Since the field $(\sum_r\Phi^r)$ decouples, the coupling $(K-ng)$ is
unrenormalized (at any order in perturbation theory). Thus~:
$\beta_K = n \beta_g$. 
Setting $n=0$ as required by the replica trick, we get~:
\debut
\beta_\sig &=& 2\({\frac{K-K_c}{K}~}\) \sig - ~2\sig^2 +\cdots
\label{betasg}\\
\beta_g &=& \inv{2}\sig^2 +\cdots\non\\
\beta_K&=& 0 \non 
\fin
So, $K$ is unrenormalized at $n=0$.
It may appears surprising that the coupling $g$ is 
renormalized although we know exactly the $g$-dependence
of any correlation functions.  We will later see
that this is not in contradiction with the renormalization group
equations.

  From equation (\ref{betasg}), we immediatly see
that in the low temperature phase $(K>K_c)$, the beta function
$\beta_\sig$ possesses a non trivial infrared zero at $\sig_*$~:
\debut
\sig_*=  \({\frac{K-K_c}{K_c}}\) + \cdots 
\quad ~~~~~~ for~~ \({\frac{K-K_c}{K_c}}\) \ll 1.
\label{IRpoint}
\fin
Notice the fact the other beta function $\beta_g$ does not
vanish at $\sig_*$~:
\debut
\beta_g(\sig_*)= \inv{2}\({\frac{K-K_c}{K_c}}\)^2 +\cdots 
\label{betastar}
\fin
Hence, even at $\sig_*$ the coupling $g$
will continue to flow. We may characterize
such pseudo-fixed point as a ``run away fixed point".
This is a particularity of the model
which has direct consequences on the correlation functions.

\bigskip
$\bullet$) {\bf Quenched correlation functions.}

Let us now analyse the renormalization group equations.
As we already said, the field $A_\mu$ would have been
generated at one loop if not present initially.
But on other hand we know that the dependence on $g$
can be completely disentangled. This apparent conflict
should have an effect on the renormalization group
equations.

Consider the correlation functions 
$G_{\al_1,\cdots}(x_1,\cdots|g,\sig)$  defined in eq.(\ref{vertex}).
They satisfy the renormalization group equations~:
\debut
\[{ \sum_p x_p^\nu\frac{\d}{\d x_p^\nu} + \sum_p \ga_p(g,\sig) -
\beta_\sig(\sig)\frac{\d}{\d\sig} - \beta_g(\sig)\frac{\d}{\d g} }\]
G_{\al_1,\cdots}(x_1,\cdots|g,\sig)=0 \non
\fin
where $\ga_p(g,\sig)$ are the anomalous dimensions.
But the $g$-dependence is explicitly known, therefore
using eq.(\ref{gdepend}) we get~:
\debut 
\[{ \sum_p x_p^\nu\frac{\d}{\d x_p^\nu} + \sum_p \ga_p(g,\sig) - 
\beta_\sig(\sig)\frac{\d}{\d\sig} - 
\beta_g(\sig) \sum_{p<q} \frac{\al_p\al_q}{K^2}\log(|x_p-x_q|^2)}\] 
G_{\al_1,\cdots}(x_1,\cdots|g,\sig)=0 \non
\fin 
Note that it is now possible to set $g$ equal to zero.
In particular at the infrared fixed
point $\sig_*$, in which $\beta_\sig(\sig_*)=0$, we get~:
\debut
\[{ \sum_p x_p^\nu\frac{\d}{\d x_p^\nu} + \sum_p \ga_p(\sig_*) - 
\beta_g(\sig_*) \sum_{p<q} \frac{\al_p\al_q}{K^2}\log(|x_p-x_q|^2)}\]
G^*_{\al_1,\cdots}(x_1,\cdots|g,\sig)=0 \label{rgeq2}
\fin 
with $\beta_g(\sig_*)$ given in eq.(\ref{betastar}).
The effect of the $g$-flow is to add the extra logarithmic term in
the renormalization group equations (\ref{rgeq2}).

This can be used to compute two-point functions at the infrared
fixed point. Consider,
\debut
G_1(x)&=& \bar { \vev{\exp\({i\al(\Phi(x)-\Phi(0)}\)}  }_{\sig_*} \non\\
G_2(x)&=& \bar {\vev{\exp\({i\al\Phi(x)}\)}
\vev{\exp\({-i\al\Phi(0)}\)}}_{\sig_*}\non
\fin
Let $\ga_{1,2}(\sig_*)$ be their anomalous dimensions at the infrared fixed point.
The renormalization group equation (\ref{rgeq2}) gives~:
\debut
G_{1,2}(x)= |x|^{-2\ga_{1,2}(\sig_*)}~
\exp\({-\frac{ \al^2\beta_g(\sig_*) }{2K^2} (\log|x|)^2 }\) 
\label{corrlog}
\fin
Notice the $(\log|x|)^2$ correction which arises from the renormalization of $g$.
This term is independent of the anomalous
dimension.

The anomalous dimensions are even in $\al$ and vanishes at $\al=0$,
therefore~:
\debut
 \ga_{1,2}(\sig_*) = \frac{\al^2}{K} \rho_{1,2}(\sig_*) + \CO(\al^4),
\quad with\quad \rho_{1}(\sig_*)=1 +\CO\({\frac{K-K_c}{K_c}}\)
\non
\fin
Expanding in power of $\al^2$, 
gives the two point functions of $\Phi$~:
\debut
 \bar {\vev{~[\Phi(x)-\Phi(0)]^2~}} =
\frac{2\rho_{1}(\sig_*)}{K}\log|x| + \frac{\beta_g(\sig_*)}{2K^2} (\log|x|)^2
\label{ver1}\\ 
 \bar {\[{~ \vev{\Phi(x)-\Phi(0)}~ }\]^2 }= 
\frac{2\rho_{2}(\sig_*)}{K}\log|x| + \frac{\beta_g(\sig_*)}{2K^2} (\log|x|)^2
\label{ver2} 
\fin
Note that the $(\log|x|)^2$ cancels in the connected correlation
function $\bar {\vev{[\Phi(x)-\Phi(0)]^2}_{conn}}$ as it should be,
since this connected correlation function is unaffected by the disorder.

There is a crossover from a $(\log|x|)$ to a $(\log|x|)^2$ behavior.
For $|x|<R_{cross}$ we have a usual $(\log|x|)$ behavior, while
for $|x|>R_{cross}$ we have a $(\log|x|)^2$ behavior.
The crossover length is $R_{cross}$ with
\debut
\log R_{cross} \sim \frac{2K\rho_{1}(\sig_*)}{\beta_g(\sig_*)}
\sim \frac{K_c^2}{(K-K_c)^2},\qquad for\quad (K-K_c)\ll 1. \non
\fin
$R_{cross}$ is exponentially large close to the phase
transition. However eqs.(\ref{ver1},\ref{ver2}) are true to all order
in perturbation theory if we can rely on the renormalization group in the 
replica symmetric approach.

We can use eqs.(\ref{ver1},\ref{ver2}) to find an estimate of the 
width of the surface for a system of finite length $L$, ie.
to find an estimate of $\bar {\vev{\Phi(0)^2}_L}$. 
We define it as the integral of the Fourier transform $\Ga(q)$
of (\ref{ver1}) using $1/L$ as ultraviolet cutoff.
At short momenta $q\ll 1/R_{cross}$, the propagator $\Ga(q)$
is dominated by the Fourier transform of the $(\log|x|)^2$
term; i.e. $\Ga(q)\sim \beta_g(\sig_*) \({\frac{\log q^2}{q^2}}\)$.
Thus~:
\begin{eqnarray} 
\bar {Q(L)}=\bar {\vev{\Phi(0)^2}_L} = \int_{1/L} d^2q \Ga(q) 
 \sim \beta_g(\sig_*) \({\log L}\)^2 \label{qbar}
\end{eqnarray}
for $L\gg R_{cross}$.
This has to be compare with the pure case which
gives a $(\inv{K}\log L^2)$ behavior.

\bigskip
$\bullet)$ {\bf A large N model.}

The formula (\ref{ver1}, \ref{ver2}) are still controversial,
theoretically as well as numerically. Variational
approaches, (part of which we will describe below),
 predict  a $(\log |x|)$ behavior
\cite{Kor,Gia,Orl}. The RG flows was also found to be unstable again
assymmetric replica perturbations \cite{Dou,Kie}.
The numerical verifications of (\ref{ver1}) are also not settled~:
a $(\log|x|)$ behavior was found in ref.\cite{Bat,Rie}, while more
recent simulations \cite{Mar} seem to indicate a $(\log|x|)^2$ behavior.
In view of this conflict, and since the variational approaches
are argued to be exact for systems with a large number of components
\cite{Mez}, in ref.\cite{BaBe}
we studied a large $N$ version of the model (\ref{actA}).
Using  RG computations based on the  (a priori symmetric) replica trick,
we find that our large $N$ model possesses a non-trivial
infrared fixed point in which the correlation functions have
the form (\ref{ver1}, \ref{ver2}) but with the $(\log|x|)^2$
term suppressed by a factor
$(1/N^3)$ compared to the $(\log|x|)$ term. 

To introduce the large $N$ version of (\ref{actA}), it is convenient
to fermionize it. The fermionic form of
the random phase sine-Gordon model is a massless Thirring model
coupled to a quenched potential $A_\mu$ and a random phase $\xi$.
To define its large $N$ version, we need to introduce $N$ Dirac
fermions with components $\psi_\pm^k$ and $\bar \psi_\pm^k$ with
$k=1,\cdots,N$. Let $z=x+iy$ and $\zb= x-iy$ be the complex coordinates
on the plane. The action is~:
\debut
S^{(N)}&=&  \int \frac{d^2x}{\pi}\({ \sum_{k=1}^N
(\psi_{-;k}\d_\zb \psi_+^k + \bar \psi_{-;k}\d_z \bar \psi_+^k)
- \frac{a}{N} (\sum_{k=1}^N \psi_{-;k}\psi_+^k)(\sum_{k=1}^N\bar \psi_{-;k}
\bar \psi_+^k) }\) \label{actferm}\\
&~&- \int \frac{d^2x}{\pi}\({ A_\zb (\sum_{k=1}^N \psi_{-;k}\psi_+^k)
+ A_z (\sum_{k=1}^N\bar \psi_{-;k} \bar \psi_+^k)
+ \xi (\sum_{k=1}^N\bar \psi_{-;k} \psi_+^k)
+ \xi^* (\sum_{k=1}^N\psi_{-;k} \bar\psi_+^k) }\) \non
\fin
In absence of disorder, it is conformally invariant.
The random potential $A_\mu=(A_\zb,A_z)$ is coupled to the
$U(1)$ currents $J_z=\sum_{k=1}^N \psi_{-;k}\psi_+^k$
and $\bar J_\zb=\sum_{k=1}^N\bar \psi_{-;k} \bar \psi_+^k$
of the unperturbed theory. At $\xi=0$, the random potential
does not break conformal invariance.

There are a priori many ways to generalize the action (\ref{actA})
to a large $N$ version. The action (\ref{actferm}) has been
designed in such way as (i) to keep the number of disordered variables
fixed, (ii) to preserve the exact conformal invariance in
absence of disorder, and (iii) to parallel as much as
possible standard properties
of large $N$ models.

\def\bbbox{\begin{picture}(3,3)(-3,-3)
\put(-3,-3){\framebox(3,3)} \end{picture}}

The fermionic action (\ref{actferm}) can be bosonized back using
non-Abelian bosonization \cite{Wit}. As usual, since the pure system describes
$N$ Dirac fermions, the pure bosonized theory will be described by
a $su(N)$ Wess-Zumino-Witten (WZW) model at level one plus a massless
free field. The $su(N)$ WZW model at level one possesses primary
fields taking values in the $(N-1)$ fundamental representations of $su(N)$.
Let $\phi_{\bbbox}^k$ and $\phi_{\bar {\bbbox} ;k}$ be
the chiral WZW primary fields which take values in the
defining representation of $su(N)$ and in its complex conjugate.
Their conformal weights are both equal to $\({\frac{N-1}{2N}}\)$.
Let us denote by $\varphi$ the gaussian free field.
The original fermions $\psi_\pm^k$  can be written as the product
of these WZW primary fields
by a vertex operator of the gaussian model. Namely,
$\psi_+^k= \phi_{\bbbox}^k~e^{\frac{i}{\sqrt{N}}\varphi}$ and
$\psi_{-;k}= \phi_{\bar {\bbbox};k}~ e^{-\frac{i}{\sqrt{N}}\varphi}$,
and similarly for the other chiral components $\bar \psi_+^k$
and $\bar \psi_{-;k}$.
The bosonic form of the action (\ref{actferm}) is~:
\debut
S^{(N)} &=& S_{wzw} + \frac{K}{2}
 \int \frac{d^2x}{4\pi} (\d_\nu\varphi)^2 \non\\
 &~&  - \int \frac{d^2x}{4\pi} \({ A_\nu(x) \d_\nu\varphi +
\xi(x) \Phi ~e^{\frac{i}{\sqrt{N}}\varphi}
+\xi^*(x) \Phi^* ~e^{-\frac{i}{\sqrt{N}}\varphi} }\) \label{actbos}
\fin
where $S_{wzw}$ refers to the $su(N)_1$ WZW action.
We have introduced the composite fields
 $\Phi(z,\zb)=\sum_{k=1}^N \phi_{\bbbox}^k(z)
 \bar {\phi_{\bar {\bbbox};k}}(\zb)$
and $\Phi^*=\sum_{k=1}^N \phi_{\bar {\bbbox};k}(z)
 \bar {\phi_{\bbbox}^k}(\zb)$.
Equivalently, $\Phi=tr_{\bbbox}(G)$ with $G$
the group valued field of the WZW model.
For $N=1$ the WZW terms are absent and we recover the  action
(\ref{actA}) of the random phase sine-Gordon model.

The renormalization group computation done in ref.\cite{BaBe} 
predicts a non-trivial infrared fixed point in the low temperature
phase with correlation function of the form (\ref{ver1}) but
with $\beta_g(\sig_*)$ given by~:
\debut
\beta_g(\sig_*) = \inv{8N^3}~ \({\frac{K-K_c}{K_c}}\)^2 + \cdots,
\quad for N \gg 1,\quad \frac{K-K_c}{K_c}\ll 1 \non
\fin
The occurence of this factor $1/N^3$ could explain
why the $(\log|x|)^2$ term does not manifest itself
in the variational approaches.

\section{The variational method~: 
the random phase sine-Gordon model.}

Various variational approaches to disorder systems have been
proposed. They can be applied after or
before disorder averaging. In the first case,
one first averages over the disorder using the replica
trick, and then implements a variational method on the
replicated model, cf.  eg \cite{Mez}. In the second case,
 one  applies a variational
method  on any sample of fixed disorder,  and then averages
over the disorder. We will present the second method
using the random phase sine-Gordon model as example, following
ref.\cite{Orl} . But it can clearly be applied to other models.
The variational method at fixed disorder leads to an
exact bound to the averaged free energy.

Consider the partition function
$Z[\xi(x)]=\int D\Phi \exp(- S(\Phi|\xi))$, where $S(\Phi|\xi)$
is the action (\ref{actA}) at $A_\mu=0$, which we recall~:
\debut
S(\Phi|\xi)=\int \frac{d^2x}{4\pi}\({ \frac{K}{2}
(\d_\nu\Phi)^2 - \xi(x) e^{i\Phi}-\xi^*(x)e^{-i\Phi}  }\)\non
\fin
At fixed disorder, we approximate the partition function
by a gaussian action~:
\debut
S_0(\Phi|G) = \half\int d^2x d^2y~
 \Phi(x) G^{-1}(x-y) \Phi(y) 
 = \half \int \frac{d^2k}{(2\pi)^2}~
\hat \Phi(k) \hat G^{-1}(k) \hat \Phi(k) \label{ansatz}
\fin
Here the hatted quantities refer to the Fourier transforms.
Notice that the ansatz (\ref{ansatz}) is choosen to be translation 
invariant, and that the gaussian is centered around the origin.
We choose a gaussian anstaz otherwise   the computations are undoable.
The kernel $G(x-y)$ is the variational parameter.

The approximated partition function is $Z_0=\int D\Phi \exp(-S_0)$.
Let $F$ and $F_0$ be the respective free energies, ie. 
$Z=e^{-F}$ and $Z_0=e^{-F_0}$. For any realization of
the disorder and for any choice of $S_0$, the following
inequality holds~:
\debut
F \leq F_0 + \vev{ \(S-S_0\)}_0  \label{convex}
\fin
where $\vev{\cdots}_0$ refers to the expectation values with 
the measure $S_0$. This is proved using the
following inequality~:
\debut
\frac{Z}{Z_0} = \vev{ e^{-(S-S_0)}}_0 \geq e^{-\vev{ \(S-S_0\)}_0}
\non
\fin
Therefore, the best approximated action $S_0$
will be find by minimazing $\({ F_0 + \vev{ \(S-S_0\)}_0 }\)$.
That is~:
\debut
\frac{\delta}{\delta \hat G(k)} 
\Bigl({ F_0 + \vev{ \(S-S_0\)}_0 }\Bigr)=0.
\label{varia}
\fin
The inequality (\ref{convex}) gives a upper bound to the free energy. 

Let us apply this general setup to our example.
As it is formulated
the method is more appropriate to study the free energy
than the correlation functions. Thus
we consider the system in a box of finite volume 
of size $Vol.=L^2$ and look for the $L$-dependence of the
free energy. That is, we use the variational free energy to 
analyse the finite size effects.

Since $S_0$ is Gaussian, $F_0$ and $\vev{ \(S-S_0\)}_0$ are
easily computed. The free energy $F_0$ per unit of volume
 is given by the logarithm of a determinant~:
\debut
\frac{F_0}{Vol.} 
=  -\half \int d^2k ~\log \hat G(k) \non
\fin
while for the expectation value of $(S-S_0)$ per unit of volume
we have~:
\debut
\frac{ \vev{\(S-S_0\)}_0}{Vol.} = \frac{\pi K}{2} 
\int \frac{d^2k}{(2\pi)^2} ~ k^2 ~\hat G(k)
- \frac{e^{-Q/2}}{Vol.}  \int \frac{d^2x}{4\pi} 
(\xi(x)+\xi^*(x)) \label{S0}
\fin
with 
\debut
Q= \vev{\Phi^2(0)}= G(0) =  
\int \frac{d^2k}{(2\pi)^2} \hat G(k) \label{Qpara}
\fin
This parameter is naturally interpreted as the width
of the system. 

Notice that the expectation value (\ref{S0}) only depends on
a particular moment of $\xi(x)$. It does not depend on
all the details of the disorder configuration. It is this particular fact
which makes the variational approach doable.

The variational equations (\ref{varia}) are simple to
compute. They determine the kernel $\hat G(k)$~:
\debut
\hat G(k) = \frac{4\pi}{K} \({\inv{k^2 + M^2} }\) \label{Gk}
\fin
where the effective mass is~:
\debut
M^2= \inv{\pi K}~\frac{e^{-Q/2}}{(Vol.)^{1/2}} ~ D_0 
\quad with \quad
D_0 = \inv{(Vol.)^{1/2}}
\int \frac{d^2x}{4\pi} (\xi(x)+\xi^*(x))  \label{meffec}
\fin
We have introduced the prefactor $(Vol.)^{1/2}$ 
in the definition of $D_0$ to make it scale invariant.
Since the effective mass $M^2$ depends on $Q$, which is a
functional of the kernel $\hat G(k)$, eqs. (\ref{Qpara},
\ref{Gk}) form a set of non linear coupled equations for
$\hat G(k)$ which we rewrite below~:
\debut
Q=\frac{4\pi}{ K} \int \frac{d^2k}{(2\pi)^2}  \({\inv{k^2 + M^2} }\)
\quad with \quad M^2= \inv{\pi K}~\frac{e^{-Q/2}}{(Vol.)^{1/2}}~D_0
\label{eqclef}
\fin

The effective mass depends on the disorder through 
its moment $D_0$. This moment can be either positive
or negative, with a symmetric probability distribution
around the origin. Thus we have to study separately
the two cases~: $D_0>0$ or $D_0<0$.
This analysis was done in ref.\cite{Orl}. 

For $D_0>0$, the effective mass is real and the Green
function $\hat G(k)$ has no pole. Therefore,
for $Q$ we get~:
\debut
Q = \int_{1/L}^{1/a} \frac{d^2k}{(2\pi)^2} \hat G(k) 
= \inv{K} \log\({\frac{a^{-2}+M^2}{L^{-2}+M^2} }\)
\label{Q1}
\fin
where $k_{UV}=1/a$ and $k_{IR}=1/L$ are the ultraviolet and
infrared cutoff. Recall that the volume is $Vol.=L^2$.
Inserting this expression in the definition (\ref{meffec})
of the effective mass gives~:
\debut
M^2= \frac{D_0}{\pi K} \inv{L}
\({\frac{a^{-2}+M^2}{L^{-2}+M^2} }\)^{-1/2K}
\label{M1}
\fin
The effective mass vanishes as $L\to\infty$, but with
different power of $L$ for $K<K_c=1$ or $K>K_c=1$.
These differences arise from the different behavior of
the last term in eq.(\ref{M1}).
For $K<K_c$, we have $M^2\ll (1/L)^2$, and therefore the
last term in eq.(\ref{M1}) behaves like $(1/L)^{1/K}$.
For $K>K_c$, the mass term dominates, $M^2\gg (1/L)^2$,
and therefore, the last term in (\ref{M1}) behaves
like $M^{1/K}$. Hence, for $D_0>0$ we obtain~:
\debut
M^2(L)\sim \(\inv{L}\)^{1+\inv{K}}\quad;\quad
Q(L)\sim \inv{K}\log L^2  \quad &~&for\quad K<K_c \label{des1}\\
(M^2(L))^{1-\inv{2K}}\sim \(\inv{L}\)\quad;\quad
Q(L)\sim \inv{2K-1}\log L^2  \quad &~&for \quad K>K_c \label{des2}
\fin

For $D_0<0$, since the situation is quite different
since the effective mass square is negative.
The Green function $\hat G(k)$ now possesses
a pole at $k^2=-M^2$. In order to analyse the effect
of this pole we have to remember that for a system in a
box of length $L$, the momenta are quantized to
discret values~: $(k_x,k_y)=(\frac{2\pi n_x}{L}, 
\frac{2\pi n_y}{L})$, with $(n_x,n_y)$ integers.
So in the key equation (\ref{eqclef}) the integral
is actually a discret sum~: $\int \frac{d^2k}{(2\pi)^2}
\to \inv{L^2} \sum_{n_x,n_y}$.
In eq. (\ref{eqclef}), we seperate the first terms
which correspond to $(n_x=\pm,n_y=\pm)$ from the others
which we approximate by an integral. We obtain~:
\debut
 Q(L) &=& \frac{4\pi}{ K L^2}\frac{4}{\({\frac{2\pi}{L}}\)^2+M^2} + \frac{4\pi}{K}
\int_{1/L}^{1/a} \frac{d^2k}{(2\pi)^2} 
\({\inv{k^2 + M^2}}\) \non\\
&=& \frac{4\pi}{ K L^2} \frac{4}{\({\frac{2\pi}{L}}\)^2+M^2} 
+ \inv{K} \log(L^2/a^2)\label{eqnega}
\fin
where we have neglected the $M^2$ dependence in the last integral.
Once again, the effective mass behaves differently as $L\to\infty$
for $K<K_c$ and $K>K_c$. These different behaviors are distinguished
by the relative importance between the two terms in eq.(\ref{eqnega}).
Indeed, suppose that $M^2\ll 1/L^2$. Then the first term in
(\ref{eqnega}) is irrelevent and therefore $Q(L)\sim\inv{K}\log L^2$.
However, inserting this value of $Q$ in the definition of $M^2$ as
a function of $Q$, cf eq.(\ref{eqclef}), we deduce then that
$M^2(L)\sim (1/L)^{1+\inv{K}}$. Thus consistency of the hypothesis
$M^2\ll 1/L^2$ requires $K<1$. When $M^2$ becomes of  order
 $1/L^2$, the first term in (\ref{eqnega}) dominates  and
$M^2$ remains frozen to this values. The expression
of $M^2$ as a function of $Q$ then tell us that $Q(L)\sim \log L^2$.
Summarizing, for $D_0<0$ we get~:
\debut
M^2(L)\sim \(\inv{L}\)^{1+\inv{K}}\quad;\quad
Q(L)\sim \inv{K}\log L^2  \quad &~&for\quad K<K_c \label{des1bis}\\
M^2(L)\sim \(\inv{L}\)^2\quad;\quad
Q(L)\sim \log L^2  \quad &~&for \quad K>K_c \label{des2bis}
\fin
Notice that in the high temperature phase $K<K_c$,
the behavior of $Q(L)$ is the same for $D_0$ positive or negative,
while this behavior is different in the low temperature phase.

These behaviors have been obtained at fixed disorder.
We can now average over the disorder. Since the effective
mass was only a function of $D_0$ which is symmetrically
distributed around the origin, the average value of $Q(L)$
is half of the sum of its values for $D_0$ positive and negative.
Hence, at large $L$, we have~:
\debut
\bar {Q(L)}\sim \inv{K}\log L^2  \quad &~&for\quad K<K_c \label{desmoy1}\\
\bar {Q(L)}\sim \frac{K}{2K-1} \log L^2  \quad &~&for \quad K>K_c
\label{desmoy2}
\fin
The averaged behavior at high temperature is the same  as in
absence of disorder. That is, the disorder is irrelevent in
the high temperature phase as we found in the previous section
using the replica approach. But the disorder is relevent
in the low temperature phase. However the result obtained
for $\bar {Q(L)}$ in the low temperature phase disagree with
the result obtained sing the symmetric replica trick,
cf eq.(\ref{qbar}).

This variational method gives poor results for the correlation functions.
Indeed, since we choose the Gaussian anstaz (\ref{ansatz})
to be centered around the orgin, the variational
one-point functions vanishes~: $\vev{ \Phi(x)}_0=0$.
But the connected correlation are unaffected by the
disorder since it is protected by the $U(1)$ symmetry.
Therefore, the variational two-point function is also
unaffected by the disorder, which is probably not
realistic. A more appropriate ansatz could be to choose
a gaussian action not centered around the origin. 

Also, this variational approach does not take one-loop
effect into account. This could be the origin of
the disagrement between the renormalization group
and the variational approaches.

\section{Replica symmetry breaking or not?}

In this section we very shortly describe how replica symmetry breaking
is incorporated in the renormalization group perturbative approach 
based on the replica method.
The aim is not to present all the details and subtilities
of the replica symmetry breaking, (there already exist
extensive reviews on the subject), but only to introduce
the main steps.

As example we choose the minimal conformal
models perturbed by random bond interaction. The
replica symmetric was studied in \cite{Lu2}, while
the study of the theory with replica symmetry breaking
was done in \cite{DoPic}.  The basic examples are the random bond Ising
or Potts models.

The simplest minimal model is the Ising model, whose disordered
version was studied above in the supersymmetric approach.
In the scaling limit, random bond interaction is represented
by a perturbation by the energy operator $\ep(x)$ with
a random coupling constant. In the Kac classification this 
operator is the $\Phi_{12}$ operator. Thus we consider
the random models~:
\debut
S[g(x)] = S_* + \int d^2x g(x) \Phi_{12}(x) \non
\fin
where $S_*$ represents the action of the corresponding
minimal conformal model, e.g the Ising or Potts models.
As we have seen, in the Ising model the disorder is
marginal and only induces logarithmic corrections.
The dimension of $\Phi_{12}$ in the Potts model
is $\dim(\Phi_{12})= \frac{4}{5}<1$. It is therefore 
a strictly  relevent disordered perturbation.

This models were 
studied in ref.\cite{Lu2} using perturbative renormalization
group cpmputation in the a priori symmetric replicated theory.

After replica, the effective action is~:
\debut
S_{eff} = \sum_z S^r_* + \sig \int d^2x \sum_{r\not= s} 
\Phi^r_{12}(x)\Phi^s_{12}(x) \label{potts1}
\fin
In eq.(\ref{potts1}), one chooses to restrict the sum to
$r\not=s$ since in the OPE of two $\Phi_{12}(x)$ operators,
which is given by the fusion rule $\Phi_{12}\times\Phi_{12}
=1 + \Phi_{13}$, only the identity arises with a singular term.
The compatilibity of this hypothesis with the
renormalization group has to be checked, and this is not a priori clear.

Replica symmetry breaking is incorporated in two steps \cite{RSB}.
One first promotes the coupling constant $\sig$
to a matrix $\sig_{rs}$~,
and then consider $\sig_{rs}$ at $n=0$ has a hierarchical Parisi matrix. 
This amounts to parametrize it by a diagonal element
$\title \sig$ and a function $\sig(x)$ with $x\in [n,1]$, $n\to 0$~:
\debut
\sig \to \sig_{rs} \to (\tilde \sig,\sig(x)) \label{matrix}
\fin
The multiplication law of two Parisi matrices parametrized 
by $(\tilde \sig_1,\sig_1(x))$ and $(\tilde \sig_2,\sig_2(x))$
is then defined by ~:
\debut
(\tilde \sig_1,\sig_1(x))\cdot(\tilde \sig_2,\sig_2(x))
= (\tilde h, h(x)) \label{mult}
\fin
with
\debut
h &=& \tilde \sig_1 \tilde \sig_2 - \bar {\sig_1\sig_2} 
\label{pari1}\\
h(x)&=& -n \sig_1(x)\sig_2(x) +
(\tilde \sig_1-\bar \sig_1)\sig_2(x)
+(\tilde \sig_2-\bar \sig_2)\sig_2(x) \non\\
&~& ~~~~-\int_n^x dy (\sig_1(x)-\sig_1(y))(\sig_2(x)-\sig_2(y)
\label{pari2}
\fin
where $\bar {f} =\int_n^1 dx f(x)$.

One can now study the consequences of this anstaz in
the renormalization group. This is done by first computing
the renormalization group equations with the matrix 
$\sig_{rs}$ and then implementing Parisi's ansatz.
As usual the one loop beta functions $\beta_{rs}=\dot \sig_{rs}$
are encoded in the operator product expansions. However,
non replica symmetric fixed points in the random mininal
conformal models only appear at two loops.

The two-loop beta function was computed in ref.\cite{DoPic}.
using an epsilon expansion. 
The central charge is parametrized as~:
\debut
c= 13 - 6(\al_+^2 + \al_+^{-2})\quad
with\quad \al_+^2= \frac{4}{3} - \ep \non
\fin
The case $\ep=0$ corresponds to the Ising model $c=\half$.
The Potts model corresponds to $\ep=\frac{2}{15}$.
The dimension of the $\Phi_{12}$ primary field is
then~:
\debut
\dim(\Phi_{12}) = 1 - \frac{3}{2}\ep \non
\fin
For $\ep=0$ the disorder is marginal, while for
$\ep \ll 1$ the disorder is slightly relevent.
This allows to implement an $\ep$-expansion, and
in particular to determine the non-trivial fixed 
point in an $\ep$-expansion. 
Since we assumed that the diagonal matrix elements
of $\sig_{rs}$ vanish, in Parisi's ansatz the matrix
$\sig_{rs}$ is represented only be a function $\sig(x)$.
The vanishing of the beta functions then determines
the possible fixed point function $\sig_*(x)$.
According to ref.\cite{DoPic}, the fixed point equation reads~:
\debut
3\ep \sig_*(x) - 2 \bar {\sig_*} \sig_*(x)
-\int^x_0 dy (\sig_*(x)-\sig_*(y))^2 +
\sig_*^3(x)+\bar {\sig_*^2}\sig_*(x)=0 \label{fixed}
\fin
with $\bar {\sig_*}=\int^1_0 dx {\sig_*}(x)$.

The replica symmetric solution corresponds to
$\sig_*(x)=const$~:
\debut
\sig_*(x)=const.= \frac{3}{2}\ep + \frac{9}{4}\ep^2 +\cdots
\non
\fin
The non replica symmetric solution is given by a solution
for which $\sig_*(x)$ growth linearly for $0<x<x_1$
and then remains constant for $x_1<x<1$. This is called
a one-step replica symmetry breaking. It is found by deriving
eq.(\ref{fixed}) with respect to $x$ as many times
as necessary. Explicitly,
one has~:
\debut
\sig_*(x)= \cases{ \inv{3}x, & if $0<x<x_1$,\cr
		\inv{3}x_1, & if $x_1<x<1$.\cr}
\quad with\quad x_1 =\frac{9}{2}\ep + \frac{27}{2}\ep^2+\cdots.\non
\fin
This solution is not present at one-loop.
It has been checked that it is a stable solution of eq.(\ref{fixed}).

The symmetric and non-symmetric solution can be distinguished
by analysing the anomalous dimensions at the corresponding
infrared fixed points. For example, the dimension of the energy operator
at the infrared fixed point differs in the two solutions
by two percent up to $\CO(\ep^3)$.
There is up to now no numerical evidence in favor of
the non-symmetric solution \cite{Picco}.

{\bf Acknowledgements}: It is a pleasure to thank H. Orland and M. Zirnbauer
for very useful discussions.

\section{Appendix: Renormalization group and OPE.}

In this appendix we gather standard informations about the
renormalization group in two dimension, cf eg. \cite{CaHouches}
\cite{Zamo}.

Consider partition functions and correlation functions
computed with the measure $\int D\phi ~ \exp(-S)$
with a perturbed action~:
\debut
S = S_* + \sum_i g^i \int d^2x \Phi_i(x) \label{A1}
\fin
where $\Phi_i(x)$ are relevent primary operators
of dimension $h_i$. Suppose that these 
fields satisfy the following  operator product expansion~:
\debut
\Phi_i(x)\Phi_j(y) = \frac{C_{ij}^k}{|x-y|^{h_i+h_j-h_k}}
\Phi_k(y) + \cdots \label{A2}
\fin
Then, the beta function at one loop is~:
\debut
\dot g^i = \beta^i(g) =
(2-h_i) g^i - \pi \sum_{jk}C^i_{jk} g^j g^k + \cdots
\label{A3}
\fin
No summation in the first term but summation over $j,k$ in the second.
The summation is over all the relevent fields generated by the
product operator expansion.

In the same way, if $\CO_\al$ is a set of operators with 
OPE with the perturbing field $\Phi_i$ given by the
structure constant $C^\al_{i\al'}$. Then the matrix of
anomalous dimensions $(\ga=-a\d_a \log Z)$ is~:
\debut
\ga^\al_{\al'} = h_\al \de^\al_{\al'} + 2\pi \sum_{j} C^\al_{j\al'} g^j + \cdots
\label{A4}
\fin
Note that we have : $\ga^i_j = 2\de^i_j -\d_i\beta^j$.

The renormalization group  equations are~:
\debut
\[{ \sum_a x_a^\nu\frac{\d}{\d x_a^\nu} + \sum_a \ga_a(g) -
\sum_j \beta^j(g)\frac{\d}{\d g^j} }\] 
\vev{\CO_1(x_1)\cdots \CO_P(x_P)} = 0
\label{A5}
\fin
For the two point functions, the integrated version of
the RG equation reads~:
\debut
\vev{\CO(R)\CO(0)}_{g(a)}
= \vev{\CO(a)\CO(0)}_{g(R)} \exp\({-2\int^{g(R)}_{g(a)}
d\bar g \frac{\ga_\Phi(\bar g)}{\beta(\bar g)} }\)
\label{A6}
\fin
where $\ga_\Phi$ is the $\ga$-function for $\Phi$.

\end{document}